\documentclass[conference]{IEEEtran}

\usepackage[utf8]{inputenc}
\usepackage{amsmath,amsthm,amssymb}
\usepackage{graphicx,enumitem,xcolor,graphics,epsfig,stmaryrd,dsfont}
\usepackage[numbers,sort&compress]{natbib}

\usepackage{xr}
\makeatletter
\newcommand*{\addFileDependency}[1]{
  \typeout{(#1)}
  \@addtofilelist{#1}
  \IfFileExists{#1}{}{\typeout{No file #1.}}
}
\makeatother

\newcommand*{\myexternaldocument}[1]{%
    \externaldocument{#1}%
    \addFileDependency{#1.tex}%
    \addFileDependency{#1.aux}%
}

\myexternaldocument{SI}

\newcommand{\inhib}{\relbar\mapsfromchar}

\newtheorem{assumption}{\textbf{Assumption}}
\newtheorem{claim}{\textbf{Claim}}
\newtheorem{proposition}{\textbf{Proposition}}
\newtheorem{theorem}{\textbf{Theorem}}
\newtheorem{lemma}{\textbf{Lemma}}
\newtheorem{definition}{\textbf{Definition}}

\newtheorem{remark}{\textbf{Remark}}
\newtheorem{example}{\textbf{Example}}
\newtheorem*{problem*}{\textbf{Problem Statement}}

\IEEEoverridecommandlockouts  

\title{\LARGE \bf \vspace{18pt} Robustness of networked systems to unintended interactions\\
\vspace{6pt}
with application to engineered genetic circuits
}

\author{Yili Qian and Domitilla Del Vecchio \thanks{$*$ Department of Mechanical Engineering, MIT, Cambridge, MA 02139, USA.  Emails: {\tt yiliqian@mit.edu} (Y. Qian) and {\tt ddv@mit.edu} (D. Del Vecchio). This work was supported in part by NSF-CMMI award \# 1727189.}
}

\begin{document}
\maketitle

\begin{abstract}
A networked dynamical system is composed of subsystems interconnected through prescribed interactions.
In many engineering applications, however, one subsystem can also affect others through ``unintended'' interactions that can significantly hamper the intended network's behavior. Although unintended interactions can be modeled as disturbance inputs to the subsystems, these disturbances depend on the network's states. As a consequence, a disturbance attenuation property of each isolated subsystem is, alone, insufficient to ensure that the network behavior is robust to unintended interactions.
In this paper, we provide sufficient conditions on subsystem dynamics and interaction maps, such that the network's behavior is robust to unintended interactions.
These conditions require that each subsystem  attenuates constant external disturbances, is monotone or ``near-monotone'', the unintended interaction map is monotone, and the prescribed interaction map does not contain feedback loops.
We employ this result to guide the design of resource-limited genetic circuits. 
More generally, our result provide conditions under which robustness of constituent subsystems is sufficient to guarantee robustness of the network to unintended interactions.

\end{abstract}

\section{Introduction}

\begin{figure}
    \centering
    \includegraphics[scale=0.4]{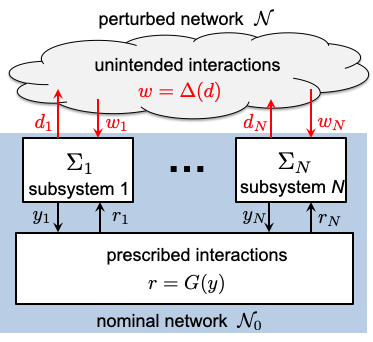}
    \caption{\textbf{Schematic of a perturbed network $\mathcal{N}$.} It is composed of $N$ subsystems interconnected via prescribed interaction map $G$ and unintended interaction map $\Delta$.
    \vspace{-10pt}}
    \label{fig:Intro}
\end{figure}

A networked system is the interconnection of input/output (I/O) subsystems through a prescribed interaction map. Many properties of networked systems can be determined using  I/O properties of the constituent subsystems and the specified interaction map \cite{Moylan1978,Fax2004,Arcak2006,Jonsson2010,Dashkovskiy2007,ZhongPingJiang2008,Rantzer2015}. 
Here, we consider the case where a networked system, which we refer to as the ``nominal network'', is perturbed by unintended interactions among subsystems (Fig.\ref{fig:Intro}).
These unintended interactions often arise from one subsystem physically perturbing the environment that comprises all other subsystems, thereby indirectly affecting their dynamics.
For example, in close formation control of aerial vehicles, the vortex created by the propulsion force of the leading vehicle can severely affect the dynamics of its neighbors, creating instability~\cite{Proud1999,Singh2000,Schumacher2000,Fierro2001}; in a wind farm with multiple turbines, the wake effect of one turbine alters the surrounding air flow, which, in turn, affects adjacent turbines, reducing efficiency~\cite{Bitar2013,Buccafusca2019}; 
in building temperature control, the temperature difference between neighboring rooms induces thermal conduction, which results in deviation of each room's temperature from its set point~\cite{Ma2012}; 
in genetic circuits, increased expression of one gene decreases the amount of resources available to express other genes, unintentionally reducing their expression levels~\cite{Gyorgy2015,Qian2017}. 

To retain the prescribed function of a network despite unintended interactions, one approach is to  co-design all subsystems and their interactions monolithically~\cite{Proud1999,Bitar2013,Buccafusca2019,Qian2017}.
A different approach, taken in networked systems research, is to allow each subsystem to be designed independent of others, thus allowing scalable network analysis and design~\cite{Jonsson2010,Fax2004,ZhongPingJiang2008,Dashkovskiy2007,Arcak2006,Siljak1972,Moylan1978,Moylan1980,Ioannou1986,Guo1999,Andreasson2014,Rantzer2015}. Specifically, work in this direction has been concerned with deriving conditions on subsystems' I/O dynamics and interaction map for network stability,
performance, and/or robustness to state-independent disturbances.
In this paper, we take the networked systems research approach. In particular, we obtain conditions for robustness to an unintended interaction map ($\Delta$ in Fig. \ref{fig:Intro}), rendering state-dependent disturbances. Our earlier work \cite{Qian2016} has studied a simplified version of this problem where the subsystems are modeled as static I/O maps.

Here, with reference to Fig. \ref{fig:Intro},
we provide mathematical conditions on the subsystems and interactions under which the behavior of the perturbed network (with unintended interactions) is arbitrarily close to that of the nominal network (without unintended interactions).
Specifically, we are interested in the network's steady state behavior and thus we define a \textit{network disturbance decoupling} (NDD) property, by which the steady state outputs from all subsystems become essentially independent of the unintended interactions. 
We prove that if (i) each constituent subsystem is monotone or near-monotone  and it can asymptotically attenuate the effect of a constant external disturbance on its output, (ii) the prescribed interactions do not contain a feedback loop, and (iii) the unintended interaction map is cooperative, then the NDD property of a network can be entirely determined by the static I/O characteristics of the subsystems.
We apply our theoretical results to guide the design of robust genetic circuits in living cells, where unintended interactions arise from resource competition and disrupt network behavior~\cite{Qian2017}. 
While solutions have appeared recently to make a single genetic subsystem robust to constant disturbances~\cite{Kelly2018,Huang2018,Aoki2019,Agrawal2018,Agrawal2019,Olsman2019}, it remains unclear the extent to which such solutions can be scaled up to enable robustness of a network of genetic subsystems to unintended interactions. 


The organization of this paper is as follows:
In Section \ref{Sec:MotivatingExample}, we present a motivating example.
In Section \ref{Sec:ProblemFormulation}, we formulate the NDD problem.
Section \ref{Sec:MonotoneSubsystems} studies networks composed of monotone subsystems and states conditions for NDD. Section \ref{Sec:SP-monotone} extends the result to non-monotone subsystems that can be reduced to a monotone system through timescale separation.
Finally, in Section \ref{Sec:Examples}, we revisit the motivating example.

\section{Motivating example} \label{Sec:MotivatingExample}
This paper is motivated by the problem of engineering robust genetic circuits (i.e., networks) in living cells~\cite{DelVecchio2016,Qian2018_ARCRAS,Hsiao2018,DelVecchio2018,McBride2019}.
These circuits allow to control the way in which a cell senses and responds to its environment, thereby offering tremendous opportunities in a number of applications, such as biomanufacturing~\cite{BioFuels}, drug delivery and therapeutics~\cite{Ruder2011}, and regenerative medicine~\cite{Johnson2017}.
Although genetic circuits have been built and used in a number of settings already, lack of robustness remains a major hurdle hampering progress~\cite{DelVecchio2018}. Among known causes of lack of robustness, competition for shared gene expression resources has appeared as a major player~\cite{Gyorgy2015,Qian2017}. In this example, we illustrate how this problem can be cast within the formulation of Fig. \ref{fig:Intro}.

A genetic circuit is composed of $N$ genetic subsystems. Each genetic subsystem contains a series of biochemical reactions that express gene $i$ to produce a protein $\mathrm{p}_i$ as output. 
In particular, the gene is first transcribed to produce mRNA $\mathrm{m}_i$ at rate $r_i$, which is then translated to produce protein $\mathrm{p}_i$ at rate $T_i$. Using $m_i$ and $p_i$ (\textit{italic}) to represent the concentrations of species $\mathrm{m}_i$ and $\mathrm{p}_i$ (roman), respectively, the state of a genetic subsystem is $x_i=[m_i,p_i]^\top$ and its output is $y_i=p_i$.
Based on mass-action kinetics, the dynamics of subsystem $i$ can be written as~\cite{bfsbook}:
\begin{align} \label{Eqn:GeneGeneral}
    &\dot{m}_i=r_i-\delta_0 m_i,&
    &\dot{p}_i=T_i(m_i)-\delta p_i,&
\end{align}
where $\delta_0$ and $\delta$ are decay rate constants of the mRNA and the protein, respectively, and $T_i(m_i)$ is the translation rate increasing with mRNA concentration $m_i$. The transcription rate of a gene $i$, $r_i$, can be modulated by the concentration of other proteins in the network, a process called \textit{transcriptional regulation}~\cite{bfsbook}. These prescribed interactions are often modeled by $r_i=G_i(y)$, where $y:=[y_1,\cdots,y_N]^\top$ and $G_i(\cdot)$ is a nonlinear function called Hill function~\cite{bfsbook}. The above descriptive framework has become standard practice to design $G$ and to tune parameters in each genetic subsystem to obtain prescribed circuit behavior, such as genetic oscillators, toggle switches, and logic gates~\cite{Elowitz2000,Gardner2000,Nielsen2016}. 

A major challenge in engineering genetic circuits is the omnipresence of unintended interactions, which severely hamper a circuit's function~\cite{Grunberg2020}.
One contributor to unintended interactions is resource competition~\cite{McBride2019}. In particular, translation of mRNA relies on the cellular resource ribosome, which is demanded by all mRNAs in the cell for translation. When mRNA $\mathrm{m}_j$ is transcribed in genetic subsystem $j$, it binds with free ribosome, reducing its availability to translate $\mathrm{m}_i$, thus unintentionally decreasing the output of subsystem $i$. Accounting for $N$ subsystems competing for a conserved pool of ribosome, the translation rate of each gene becomes~(see \cite{Qian2016} for derivation):
\begin{align} \label{Eqn:GeneComp}
    T_i=T_i(m_i,w_i)=\frac{\alpha_i\cdot(m_i/\kappa_i)}{1+m_i/\kappa_i+w_i},\;
    w_i = \sum_{j \neq i}\frac{m_j}{\kappa_j},
\end{align}
where $\alpha_i$ is the translation rate constant, $\kappa_i$ is the dissociation constant that decreases with the affinity of $\mathrm{m}_i$ with the ribosome, and $w_i$ is the ribosome demand by all  other subsystems in the circuit. Because translation rate $T_i$ decreases with $w_i$, by substituting (\ref{Eqn:GeneComp}) into (\ref{Eqn:GeneGeneral}), we observe that the  output $y_i=p_i$ now decreases with $m_j$. These create unintended interactions and give rise to unexpected circuit behavior~\cite{Qian2017}. Hence, a genetic circuit with ribosome competition can be regarded as a perturbed network with subsystem dynamics (\ref{Eqn:GeneGeneral}) with $T_i=T_i(m_i,w_i)$, with prescribed interaction (i.e., transcriptional regulation) map $G(\cdot)$, and with unintended interaction map $\Delta(\cdot)$: $w_i=\sum_{j \neq i}d_j$, where $d_i=m_i/\kappa_i$ is the disturbance output of subsystem $i$.

To reduce the dependence of each subsystem's output $y_i$ on  disturbance $w_i$, an additional molecule, called small RNA (sRNA), was introduced into each genetic subsystem to create a biomolecular feedback control mechanism~\cite{Huang2018}. 
The dynamics in such a \textit{feedback-regulated subsystem} can be described by the following mass-action kinetic model:
\begin{equation} \label{Eqn:sRNA-Dyn}
    \begin{aligned}
    \dot{m}_i &= \frac{1}{\varepsilon_i} r_i-\frac{1}{\varepsilon_i} \lambda_i m_i s_i - \delta_0 m_i,\\
    \dot{s}_i &= \frac{1}{\varepsilon_i} \beta_i p_i - \frac{1}{\varepsilon_i} \lambda_i m_i s_i - \delta_0 s_i,\\
    \dot{p}_i & = T_i(m_i,w_i)-\delta p_i,
\end{aligned}
\end{equation}
where $s_i$ is the concentration of sRNA, $\lambda_i$, $\beta_i$ are constant parameters, and $\varepsilon_i$ is a small design parameter that can be decreased experimentally (see~\cite{Huang2018}). When $w_i$ is a constant, state-independent disturbance, it has been shown that the steady state output of (\ref{Eqn:sRNA-Dyn}) satisfies $\lim_{\varepsilon_i \to 0^+}y_i=r_i/\beta_i$, which is independent of $w_i$. This asymptotic static disturbance attenuation property is attained if the constant reference input takes value in an admissible set $\bar{\mathcal{R}}_i:= \lbrace 0 \leq r_i<\alpha_i/\beta_i \rbrace$~\cite{Qian2018_JRSI}. The situation $r_i \geq \alpha_i/\beta_i$ physically corresponds to a scenario where the desired output cannot be reached even with all available ribosomes translating $\mathrm{m}_i$. 

\begin{figure}
    \centering
    \includegraphics[scale=0.43]{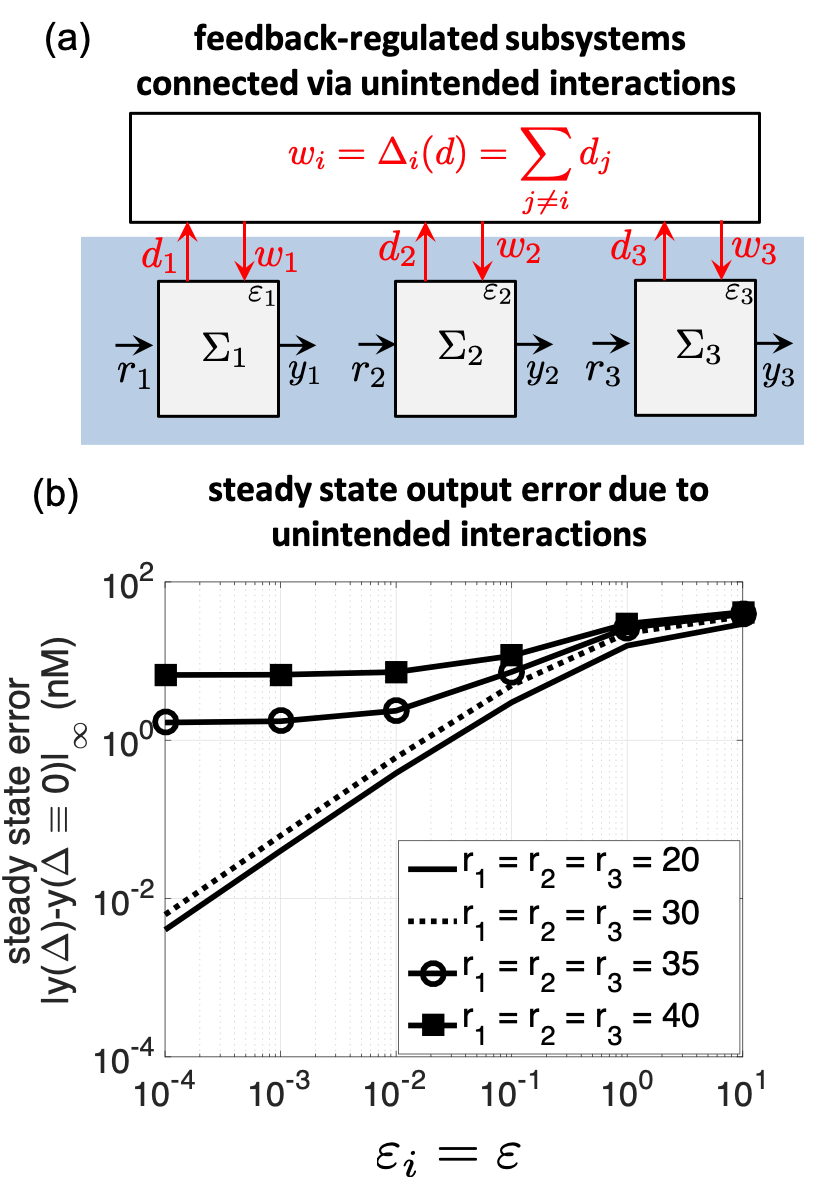}    
    \caption{\textbf{Network disturbance decoupling for feedback-regulated genetic subsystems with independent reference inputs.} (a) The nominal network $\mathcal{N}_0$ (shaded in blue) consists of three feedback-regulated genetic subsystems (\ref{Eqn:sRNA-Dyn}), each taking an independent but identical reference input $r_i=r_0$. The subsystems are coupled through unintended interactions arising from resource conservation ($w_i = \Delta_i(d)$), leading to the perturbed network $\mathcal{N}$. (b) Steady state error (vector $\infty$-norm) between the outputs of the perturbed and the nominal networks as $\varepsilon_i=\varepsilon$ varies. For every $\varepsilon_i=\varepsilon$ and $r_i=r_0$, the trajectory converges to an asymptotically stable equilibrium. Subsystems have identical parameters: $\alpha_i=100$ nM/hr, $\lambda_i=1$ $(\text{nM}\cdot \text{hr})^{-1}$, $\delta=1$ $\text{hr}^{-1}$, $\beta_i=1$ $\text{hr}^{-1}$, and $\kappa_i=1$ $\text{nM}$ for all $i$. Based on these parameters and for all $r_i$ levels chosen, we find $r_i \in \bar{R}_i$ and hence each subsystem in isolation can asymptotically attenuate a constant disturbance as $\varepsilon_i$ decreases. \vspace{-10pt}}
    \label{fig:IndpInput}
\end{figure}

Given that each subsystem can asymptotically reject disturbance $w_i$  to reach set-point $r_i/\beta_i$, it is tempting to use multiple such feedback controllers, one in each genetic subsystem, to ensure that the output of multiple feedback-regulated subsystems become independent of $w_i$, that is, of ribosome usage. This approach, however, can fail depending on the value of reference input $r_i$ to each subsystem. Specifically, we simulated the network in Fig.\ref{fig:IndpInput}a, which is composed of 3 feedback-regulated genetic subsystems with the dynamics in (\ref{Eqn:sRNA-Dyn}) but no prescribed interactions among them (i.e., $r_i(t) \equiv r_0$ for all $i$). 
We chose simulation parameters such that $r_i \in \bar{\mathcal{R}}_i$, hence each subsystem in isolation can asymptotically reject any constant disturbance as $\varepsilon_i$ is decreased. 
However, as shown in Fig.\ref{fig:IndpInput}b, we found that decreasing $\varepsilon_i$ for all subsystems fails to decrease the tracking error for large reference input values despite $r_i \in \bar{\mathcal{R}}_i$.

These simulation results demonstrate that even if all constituent subsystems of a network can attenuate constant, state-independent disturbances in isolation, this robustness property may be lost when disturbances are state-dependent through an unintended interaction map $w=\Delta(d)$.
Specifically, in this case, the problem occurs because $d_i$ reflects the ``control effort'' of the feedback regulation mechanism in subsystem $i$.
Hence, when $\varepsilon_i \to 0^+$ to improve disturbance attenuation  of subsystem $i$, depending on $r_i$ level, disturbance output $d_i$ may grow unbounded, leading to $w_j \to \infty$, which cannot be compensated by the control effort in subsystem $j$.
The result in this paper allows us to place sufficient conditions on subsystem dynamics, $\Delta$, and $G$ such that this problem does not occur.\vspace{-4pt}

\section{Problem formulation} \label{Sec:ProblemFormulation}
After introducing some notations, we present our system setup. Specifically, We describe mathematical conditions that restrict the class of subsystems we consider. We then formally define the NDD problem.\\

\noindent \textbf{Notations:} For a vector $v \in \mathbb{R}^n$, we denote $|v|:=\max_i|v_i|$ for vector $\infty$-norm. For a signal $v(t):\mathbb{R} \to \mathbb{R}^n$, its $\infty$-norm is denoted as $\|v\|:= \sup_{t \geq 0}|v(t)|$.
For a closed set $\mathcal{A}$ and a vector $x$, $\text{dist} \lbrace x,\mathcal{A} \rbrace = \min_{s \in \mathcal{A}}|x-s|$. For a time-dependent function $x(t)$, we will use the following notations:
\begin{align*}
    &\lim_{t \to \infty}\text{dist} \lbrace x(t),\mathcal{A}\rbrace=0&
    &\Leftrightarrow&
    &x(t) \to \mathcal{A},&
    \\
    &\lim_{t \to \infty}\text{dist} \lbrace x(t),\mathcal{A}\rbrace \leq \mu&
    &\Leftrightarrow&
    &x(t) \xrightarrow{\mu} \mathcal{A}.&
\end{align*}
The comparison operators $<$, $\leq$, as well as $\min$ and $\max$ operations are defined component-wise. 
The set $[a,b] := \lbrace x \in \mathbb{R}^n: a \leq x \leq b\rbrace$, where $a \leq b$, defines a box in $\mathbb{R}^n$. 
Concatenation of $N$ $a$-dimensional vectors $x_1,\cdots,x_N$ is written as $x:=[x_1^\top,\cdots,x_N^\top]^\top \in \mathbb{R}^{aN}$. Similarly, given $N$ vector-valued functions $f_1(x_1),\cdots,f_N(x_N)$ with $f_i:\mathbb{R}^a \to \mathbb{R}^b$ for all $i$, we write the stacked function as $f(x):=[f_1^\top(x_1),\cdots,f_N^\top(x_N)]^\top: \mathbb{R}^{aN} \to \mathbb{R}^{bN}$. For sets $\mathcal{A}_1,\cdots,\mathcal{A}_N$, we write $\mathcal{A}:=\prod_{i=1}^N\mathcal{A}_i$. A scalar continuous function $\alpha(x)$ with $\alpha(0)=0$ is of class $\mathcal{K}_0$ ($\mathcal{K}$) if it is non-decreasing (strictly increasing) with $x$. For a $n \times m$ matrix $A$, $\text{sign}(A)_{ij}=1$ if $A_{ij} \geq 0$ and $\text{sign}(A_{ij})=-1$  otherwise. 
A function $f(x,y)$ is said to be Lipschitz continuous in $x \in \mathcal{X}$ uniformly in $y \in \mathcal{Y}$ if there exists a constant $L>0$ such that for all $y \in \mathcal{Y}$, $|f(x^+,y)-f(x^-,y)| \leq L|x^+-x^-|$ for any $x^-,x^+ \in \bar{\mathcal{X}}$. 
\hfill $\triangledown \vspace{10pt}$

With reference to Fig.\ref{fig:Intro}, a \textit{perturbed network} $\mathcal{N}$ is a tuple $(\Sigma,G,\Delta)$, where $\Sigma:=(\Sigma_1,\cdots,\Sigma_N)$ is a set of $N$ subsystems, and $G$ and $\Delta$ describe the prescribed and unintended interaction maps, respectively. Each subsystem $\Sigma_i=\Sigma_i(\varepsilon_i)$ is parameterized by a positive parameter $\varepsilon_i$ and follows the dynamics:
\begin{align} \label{Eqn:Setup.Subsys}
    &\dot{x}_i=f_i(x_i,r_i,w_i;\varepsilon_i),&
    &y_i=l_i(x_i),&
    &d_i=\rho_i(x_i),&
\end{align}
where $x_i$ is the state variable evolving in $\mathcal{X}_i \subseteq \mathbb{R}^n$.
Signals $r_i$ and $w_i$ are reference and disturbance inputs, respectively, taking values on sets $\mathcal{R}_i$ and $\mathcal{W}_i$ that contain the origin; $y_i$ and $d_i$ are prescribed and disturbance outputs, respectively, taking values on $\mathcal{Y}_i$ and $\mathcal{D}_i$. 
For each fixed $\varepsilon_i$, we assume the function $f_i$ is differentiable and locally Lipschitz on $\mathcal{X}_i \times \mathcal{R}_i \times \mathcal{W}_i$. 
The output functions
$l_i,\rho_i$ are assumed to be differentiable and locally Lipschitz on $\mathcal{X}_i$.
For simplicity, we consider I/O signals $r_i,w_i,y_i$ and $d_i$ to be scalars, and write $u_i:=[r_i,w_i]^\top$ and $q_i:=[y_i,d_i]^\top$. Because of this, with slight abuse of notation, for any function $f(\cdot)$ with vector argument $u_i=[r_i,w_i]^\top$, the notation $f(u_i)$ is used interchangeably with $f(r_i,w_i)$ for convenience.
\begin{assumption} \label{Ass:Setup.SubsysStability}
\normalfont (Subsystem stability). There exists $\varepsilon_i^*>0$ such that for each fixed $(r_i,w_i) \in \mathcal{R}_i \times \mathcal{W}_i$ and $0<\varepsilon_i \leq \varepsilon_i^*$, system (\ref{Eqn:Setup.Subsys}) has a globally asymptotically stable (GAS) equilibrium $\varphi_i(r_i,w_i;\varepsilon_i)$, that is, for all initial conditions $x_i^0\in\mathcal{X}_i$, $\lim_{t \to \infty}x_i(t,r_i,w_i;\varepsilon_i)=\varphi_i(r_i,w_i;\varepsilon_i)$.
\hfill $\triangledown$
\end{assumption}
\noindent If Assumption \ref{Ass:Setup.SubsysStability} is satisfied, $\varphi_i(\cdot,\cdot;\varepsilon_i)$ is called the \textit{static input/state (I/S) characteristic} of $\Sigma_i$. 
The corresponding \textit{static I/O characteristic} for the prescribed output is:
\begin{align} 
    y_i&=h_i(r_i,w_i;\varepsilon_i):=l_i\circ \varphi_i(r_i,w_i;\varepsilon_i).\label{Eqn:Setup.refoutput}
\end{align}

\begin{assumption} \label{Ass:Setup.DistAttn}
\normalfont (Subsystem disturbance attenuation). There exists class $\mathcal{K}$ functions $\alpha_i(\cdot)$ and $\alpha_i^0(\cdot)$, a non-empty compact set $\bar{\mathcal{R}}_i \subseteq \mathcal{R}_i$, a constant $\varepsilon_i^*>0$, and a bounded function $H_i(r_i)$ such that 
\begin{align} \label{Eqn:Setup.DistAttn}
    |h_i(r_i,w_i;\varepsilon_i)-H_i(r_i)| \leq \alpha_i(\varepsilon_i)|w_i| + \alpha_i^0(\varepsilon_i)
\end{align}
for every fixed $(r_i,w_i) \in \bar{\mathcal{R}}_i \times \mathcal{W}_i$ and $0<\varepsilon_i \leq \varepsilon_i^*$. 
\hfill $\triangledown$
\end{assumption}
\noindent We call $H_i(r_i)$ the \textit{nominal static I/O characteristic} because it is independent of $w_i$. 
According to Assumption \ref{Ass:Setup.DistAttn}, for any bounded and fixed disturbance input $w_i$, the steady state prescribed output $y_i=h_i(r_i,w_i;\varepsilon_i)$ deviates at most $\mathcal{O}(\varepsilon_i)$ from $H_i(r_i)$. 
The set $\bar{\mathcal{R}}_i$ is the \textit{admissible reference input set}, where  (\ref{Eqn:Setup.DistAttn}) holds.

The subsystems are connected through a  static intended interaction map
\begin{align} \label{Eqn:Setup.Intended}
    r=G(y).
\end{align}
\noindent 
In a perturbed network, the disturbance output of subsystem $i$, $d_i$, perturbs subsystem $j$ through a disturbance input $w_j$. The dependence of $w_j$ on $d_i$ gives rise to unintended interactions among subsystems, which we model using a static unintended interaction map
\begin{align} \label{Eqn:Setup.unintended}
    w=\Delta(d).
\end{align}
\noindent 
We assume that both maps $G(\cdot)$ and $\Delta(\cdot)$ are globally Lipschitz. We use $y(t;\varepsilon,\Delta)$ to represent the stacked outputs of the perturbed network consisting of (\ref{Eqn:Setup.Subsys}), (\ref{Eqn:Setup.Intended}), and (\ref{Eqn:Setup.unintended}), and write $y(t;\varepsilon,0)$ for the stacked outputs of a \textit{nominal network} $\mathcal{N}_0=(\Sigma,G,\Delta \equiv 0)$ consisting of (\ref{Eqn:Setup.Subsys}), (\ref{Eqn:Setup.Intended}), but without disturbance input (i.e., $w \equiv 0$). 

\begin{definition} \label{Def:mu-NDD}
\normalfont (NDD). Given $\mu > 0$ and a fixed $\varepsilon$, the perturbed network $\mathcal{N}(\varepsilon)=(\Sigma(\varepsilon),G,\Delta)$ is said to have the $\mu$-network disturbance decoupling ($\mu$-NDD) property if
\begin{align*}
    \limsup_{t \to \infty}|y(t;\varepsilon,\Delta)-y(t;\varepsilon,0)| \leq \mu
\end{align*}
for all initial conditions $x^0 \in \mathcal{X}$.
\hfill $\triangledown$
\end{definition}


\noindent For small $\mu$, the output of $\mathcal{N}$ becomes close to that of the nominal network $\mathcal{N}_0$. The $\mu$-NDD property therefore quantifies network robust performance with respect to the unintended interaction map $\Delta$. In general, asymptotic static disturbance attenuation of the subsystems is insufficient to guarantee $\mu$-NDD for arbitrarily small $\mu$. For example, the unintended interactions may result in $\lim_{\varepsilon \to 0^+}|w(t;\varepsilon)| \to \infty$, as shown in the motivating example of Section \ref{Sec:MotivatingExample}, or they may de-stabilize the network. 

\begin{problem*} 
\normalfont Given a perturbed network $\mathcal{N}(\varepsilon)=(\Sigma(\varepsilon),G,\Delta)$ consisting of subsystems with the asymptotic static disturbance attenuation property (\ref{Eqn:Setup.DistAttn}), determine conditions on $\Sigma_i(\varepsilon_i)$, $G$, and $\Delta$ such that given any $\mu>0$, $\mu$-NDD can be achieved if $\varepsilon_i$ is sufficiently small for every $i$.
\hfill $\triangledown$
\end{problem*}

\noindent Solution to the NDD problem identifies a class of perturbed networks that are robust to unintended interactions, in the sense that any effect arising from unintended interactions can be mitigated by simply improving disturbance attenuation of the constituent subsystems (i.e., decreasing $\varepsilon_i$).
As we  demonstrate next, one class of such networks are those with certain monotonicity properties.  

\section{Network disturbance decoupling with monotone subsystems} \label{Sec:MonotoneSubsystems}
After introducing background on monotone systems, we provide mathematical conditions to solve the NDD problem for networks composed of monotone subsystems. 

\subsection{Technical background: Monotone systems}
\label{Sec:Background}
We present some basic concepts on monotone systems theory and mixed-monotone functions. A more complete and in-depth treatment of these topics can be found in~\cite{Smith1995,Angeli2003,Angeli2014,Coogan2015,Smith2008}. 
\begin{definition} \label{Def:MixedMonotone}
\normalfont (\cite{Coogan2015}). A function $f:\mathcal{X} \to \mathcal{Y}$ is \textit{mixed-monotone} if there exists a function $\hat{f}:\mathcal{X}^2 \to \mathcal{Y}$, called a \textit{decomposition function} of $f(\cdot)$, such that for all $x,x_1,x_2,z \in \mathcal{X}$ the following are satisfied: (i) $f(x)=\hat{f}(x,x)$, (ii) $x_1 \leq x_2 \Rightarrow \hat{f}(x_1,z) \leq \hat{f}(x_2,z)$, and (iii) $x_1 \leq x_2 \Rightarrow \hat{f}(z,x_2) \leq \hat{f}(z,x_1)$. 
\hfill $\triangledown$
\end{definition}
\noindent According to the above definition, take any $x^- \leq x \leq x^+$, we have $\hat{f}(x^-,x^+) \leq f(x) \leq \hat{f}(x^+,x^-)$. 
A differentiable function $f:\mathbb{R}^m \to \mathbb{R}^n$ has \textit{sign-stable} partial derivatives if there exists a matrix $\Lambda \in \mathbb{R}^{n \times m}$, whose elements $\Lambda_{ij}$ take values in $\lbrace 1,-1 \rbrace$ and satisfy $\Lambda_{ij}(\partial f_i/\partial x_j) \geq 0$ for all $i,j$ and $x$.
If $f$ has \text{sign-stable} partial derivatives, then one decomposition function of $f$ can be found through $\Lambda$. 
In particular, let
\begin{align} \label{Eqn:SignPattern}
    &\Lambda^-=-\min(0,\Lambda),&
    &\Lambda^+=\mathbf{1}_{m \times n}-\Lambda^-,&
\end{align}
define a vector function $\hat{f}(x^+,x^-):\mathbb{R}^{2m} \to \mathbb{R}^n$ whose $i$-th element is:
\begin{align}
\label{Eqn:DecompFcnConst}
\hat{f}_i(x^+,x^-):=f_i\left(\text{diag}(\Lambda^+_{i})\cdot x^++\text{diag}(\Lambda^-_{i})\cdot x^-\right),
\end{align}
where $\Lambda_{i}^+$ (or $\Lambda_{i}^-$) is the $i$-th row of $\Lambda^+$ (or $\Lambda^-$, respectively). Then, $\hat{f}$ is a decomposition function of $f$. In particular, we call $\hat{f}$ the \textit{canonical decomposition function} of $f$.

\begin{example}
\normalfont Given a constant matrix $A$, the function $f(x)=A x$ is mixed-monotone. Its canonical decomposition function is $\hat{f}(x^+,x^-)=A^+ x^+ + A^- x^-$, where
\begin{align*}
    &A^-_{ij}:= \begin{cases}
    A_{ij},&\;\; \text{if } A_{ij} < 0,\\
    0, &\;\;\text{otherwise},
    \end{cases}&
    &A^+:= A - A^-.&
\end{align*}
\end{example}
\noindent For any $x^- \leq x \leq x^+$, it can be verified that $\hat{f}(x^-,x^+)=A^+ x^- + A^- x^+ \leq f(x)=Ax \leq A^+ x^+ + A^- x^- = \hat{f}(x^+,x^-)$. 
\hfill $\triangledown$

\begin{lemma} \label{Lemma:ComposeMixedMonotone}
\normalfont Let $f$ and $g$ be two mixed-monotone functions with decomposition functions $\hat{f}$ and $\hat{g}$, respectively. Then $h:=f \circ g$ is also mixed-monotone and $\hat{h}(x_1,x_2):=\hat{f}(\hat{g}(x_1,x_2),\hat{g}(x_2,x_1))$ is a decomposition function of $h$.
\hfill $\triangledown$
\end{lemma}
Now we consider a system with input $u(t)$ and output $q(t)$:
\begin{align} \label{Eqn:ISSys}
    &\dot{x}=f(x,u),&
    &q=L(x),&
\end{align}
where $f:\mathbb{R}^n \times \mathbb{R}^m \to \mathbb{R}^n$ and $L:\mathbb{R}^n \to \mathbb{R}^b$ are differentiable and their partial derivatives with respect to $x$ and $u$ are sign-stable. We review the notion of (orthant) input/state (I/S) monotone systems \cite{Angeli2014}. 
\begin{definition}
\normalfont (\cite{Angeli2003,Angeli2014}). System (\ref{Eqn:ISSys}) is I/S monotone if there exists vectors $\sigma^u \in \mathbb{R}^m$ and $\sigma^x \in \mathbb{R}^n$, whose elements take values in $\lbrace 1,-1\rbrace$, such that
\begin{align*} 
&\sigma^x_{i}\sigma^x_{j} \frac{\partial f_i}{\partial x_j}(x,u) \geq 0,&
&\sigma^x_{i}\sigma^u_{k} \frac{\partial f_i}{\partial u_k}(x,u) \geq 0,&
\end{align*}
for all indices $i\neq j$, $k$, and for all $x,u$.
Specifically, the system is said to be I/S monotone with respect to the partial order pair $(\sigma^u;\sigma^x)$. 
\hfill $\triangledown$
\end{definition}
\noindent If for each fixed $u$, the I/S monotone system (\ref{Eqn:ISSys}) has a GAS equilibrium $x=\varphi(u)$, then $\varphi(\cdot)$ is called the static I/S characteristic of (\ref{Eqn:ISSys}). The {static I/S characteristic} of an I/S monotone system has sign-stable partial derivatives~\cite{Angeli2003}. In particular, the sign pattern of $(\partial \varphi/\partial u)$ is $\Lambda=\sigma^x(\sigma^u)^\top$, and the canonical decomposition function of $\varphi$ can then be found according to (\ref{Eqn:DecompFcnConst}). 
An important property of I/S monotone systems is the following convergent-input-convergent-state/output property.

\begin{definition} \label{Def:CICS}
\normalfont (\cite{Angeli2003,Angeli2014}.) System (\ref{Eqn:ISSys}) is convergent-input-convergent-state if there exists a function $\phi(\cdot,\cdot):\mathbb{R}^{2m} \to \mathbb{R}^{2n}$, called an \textit{I/S gain function} of (\ref{Eqn:ISSys}), such that for any $u^-,u^+$, if $u(t) \to [u^-,u^+]$, then $x(t) \to [\phi(u^-,u^+),\phi(u^+,u^-)]$. Similarly, it is convergent-input-convergent-output if there exists a function $\psi(\cdot,\cdot):\mathbb{R}^{2m} \to \mathbb{R}^{2b}$, called an \textit{I/O gain function}, such that for any $u^-,u^+$, if $u(t) \to [u^-,u^+]$, then $q(t) \to [\psi(u^-,u^+),\psi(u^+,u^-)]$.
\hfill $\triangledown$
\end{definition}

\begin{figure}
    \centering
    \includegraphics[scale=0.42]{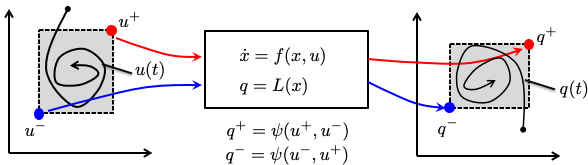}
    \caption{\textbf{A graphical representation of the I/O gain function $\psi$ for system (\ref{Eqn:ISSys})}. If the input $u(t)$ ultimately enters the box $[u^-,u^+]$, the output $q(t)$ ultimately converges to the box $[\psi(u^-,u^+),\psi(u^+,u^-)]$. This schematic assumes system (\ref{Eqn:ISSys}) is cooperative, that is, $(\sigma^u;\sigma^x)=(1;\mathbf{1}_n)$, and $\partial L/\partial x \geq 0$ for all $x$.\vspace{-10pt}}
    \label{fig:IOGain}
\end{figure}

\noindent A graphical representation of a convergent-input-convergent-output system with I/O gain function $\psi$ is shown in Fig.\ref{fig:IOGain}. 
If the input $u(t)$ eventually enters the box $[u^-,u^+]$, output $q(t)$ will eventually converge to the box $[\psi(u^-,u^+),\psi(u^+,u^-)]$. 

\begin{lemma} \label{Lemma:CICS}
\normalfont Suppose that (\ref{Eqn:ISSys}) is monotone with a static I/S characteristic $x=\varphi(u)$, then it is convergent-input-convergent-state. 
Additionally, if the output function $L(x)$ is mixed-monotone with a decomposition function $\hat{L}(x^+,x^-)$, then (\ref{Eqn:ISSys}) is convergent-input-convergent-output. Specifically, let $\hat{\varphi}$ be the canonical decomposition function of $\varphi$, then an I/O gain function of (\ref{Eqn:ISSys}) is $\psi(u^+,u^-):=\hat{L}(\hat{\varphi}(u^+,u^-),\hat{\varphi}(u^-,u^+))$.
\hfill $\triangledown$
\end{lemma}

\noindent Proof for convergence of $x(t)$ can be found in \cite{Angeli2014} (Lemma 2). Convergence of $q(t)$ is a consequence of Lemma \ref{Lemma:ComposeMixedMonotone}.
I/S monotonicity of system (\ref{Eqn:ISSys}) can be determined by simple graphical conditions~\cite{Kunze1999}. Specifically,
the \textit{incidence graph induced by} $f$ is a signed digraph. Each element in the $(n+m)$-vector $\xi=[x^\top,u^\top]^\top$ is a node. There is a directed edge $(\xi_i,x_j)$ from $\xi_i$ to $x_j$ if $\text{sign}(\partial f_{j}/\partial \xi_i) \neq 0$ for some $\xi$. Each edge $(\xi_i,x_j)$ is associated with a sign defined as $\text{sign}(\partial f_j/\partial \xi_i)$. 
An \textit{undirected cycle} is a sequence of nodes $\xi_{c_1},\cdots,\xi_{c_k}$ such that $\xi_{c_1} = \xi_{c_k}$ and for each $1 \leq i \leq (k-1)$, either edge $(\xi_{c_i},\xi_{c_{i+1}})$ exists or edge $(\xi_{c_{i+1}},\xi_{c_i})$ exists. The sign of this cycle is the product of the signs of all edges constituting the cycle.
\begin{lemma} \label{Lemma:graphical}
\normalfont (\cite{Kunze1999,Angeli2004b}). System (\ref{Eqn:ISSys}) is I/S monotone if and only if the incidence graph induced by $f$ does not contain an undirected negative cycle.
\hfill $\triangledown$
\end{lemma}

\subsection{Conditions on subsystems and interaction maps} \label{Sec:NDD-monotone.conditions}

\noindent Here we provide a set of sufficient conditions on the subsystem dynamics and prescribed/unintended interaction maps for NDD. 
These conditions are centered around the subsystems having the I/S monotonicity property, which we 
assume to hold on boxes $\mathcal{X}_i,\mathcal{R}_i,\mathcal{W}_i,\mathcal{Y}_i$, and $\mathcal{D}_i$. 
These boxes are Cartesian products of (possibly unbounded) closed real intervals. 
Additionally, we assume that for all $(r_i(t),w_i(t))$ taking values on $ \mathcal{R}_i \times \mathcal{W}_i$ and for all $\varepsilon_i>0$, the set $\mathcal{X}_i$ is positively invariant under the subsystems dynamics (\ref{Eqn:Setup.Subsys}).
For example, in biomolecular systems, $\mathcal{X}_i,\mathcal{R}_i,\mathcal{W}_i,\mathcal{Y}_i$, and $\mathcal{D}_i$ can be chosen as the non-negative orthant, because the state variables and I/O signals represent species concentrations and are thus non-negative.

\begin{assumption} \label{Ass:monotone.SubsysCooperative}
\normalfont (Subsystem monotonicity). For every $\varepsilon_i \in (0,\varepsilon_i^*]$, each subsystem $\Sigma_i$ in (\ref{Eqn:Setup.Subsys}) is I/S monotone with respect to the partial orders $(\sigma^u;\sigma^x)$. The partial derivatives of output functions $l_i$ and $\rho_i$ are sign-stable.    
\hfill $\triangledown$
\end{assumption}


Due to Assumptions \ref{Ass:Setup.SubsysStability} and \ref{Ass:monotone.SubsysCooperative}, the subsystem I/S characteristic $\varphi_i$ is mixed-monotone. Let  $\hat{\varphi}_i(u_i^+,u_i^-;\varepsilon_i)$ and $\hat{\rho}_i(x_i^+,x_i^-)$ be the canonical decomposition functions of $\varphi_i(u_i;\varepsilon_i)$ and $\rho_i(x_i)$, respectively. We follow Lemma \ref{Lemma:CICS} and define the disturbance I/O gain function of $\Sigma_i$ as:
\begin{align} \label{Eqn:monotone.IOG}
    \psi_i(u_i^+,u_i^-;\varepsilon_i):=\hat{\rho}_i(\hat{\varphi}_i(u_i^+,u_i^-;\varepsilon_i),\hat{\varphi}_i(u_i^-,u_i^+;\varepsilon_i)).
\end{align}
We assume that increasing disturbance output from $\Sigma_i$ does not decrease disturbance input to $\Sigma_j$. This is a mild assumption satisfied in many scenarios, including our motivating example in Section \ref{Sec:MotivatingExample}, as we will show in Section \ref{Sec:Examples}.
\begin{assumption}\label{Ass:UnintendedInteractions}
\normalfont (Unintended interactions). The unintended interaction map $\Delta(\cdot)$ is cooperative, that is, $\Delta_i(d_j^-) \leq \Delta_i(d_j^+)$ for all $i,j$ and $d_j^- \leq d_j^+$.
\hfill $\triangledown$
\end{assumption}

\noindent The prescribed interaction map is assumed to have a simple structure.
\begin{assumption} \label{Ass:IntendedInteractions}
\normalfont (Intended interaction). The  intended interaction map $r=G(y)$ does not contain any feedback loop, that is, $\partial G_i/\partial y_j \equiv 0$ for all $j \geq i$. 
\hfill $\triangledown$
\end{assumption}


\noindent Given Assumptions \ref{Ass:Setup.SubsysStability} and \ref{Ass:IntendedInteractions}, because $G$ does not contain feedback loops, equation $r=G \circ H(r)$ has a unique solution $r^*=[r_1^*,\cdots,r_N^*]^\top$. We call $r^*$ the \textit{nominal reference input} to the network, since $r^*$ is computed using $G$ and the subsystem nominal static I/O characteristic $y_i=H_i(r_i)$, which is independent of $w_i$. 
We use
\begin{align} \label{Eqn:monotone.fixedinputIOG}
    \psi_i^*(w_i^+,w_i^-;r_i^*,\varepsilon_i):=\psi_i(r_i^*,w_i^+,r_i^*,w_i^-;\varepsilon_i)
\end{align}
to represent a subsystem's disturbance I/O gain function for a fixed $r_i^*$.
If $w_i \to [w_i^-,w_i^+]$ and $r_i \equiv r_i^*$, then the disturbance output $d_i$ is ultimately bounded in the box $[\psi_i^*(w_i^-,w_i^+;r_i^*,\varepsilon_i),\psi_i^*(w_i^+,w_i^-;r_i^*,\varepsilon_i)]$. We will use $\psi_i^*$ to elicit conditions for NDD and use $v^\pm$ to represent vector concatenation $v^\pm:=[(v^-)^\top,(v^+)^\top]^\top$. 
Finally, we impose the following technical assumption on each subsystem's static characteristic and disturbance I/O gain function.

\begin{assumption} \label{Ass:LipschitzConditions}
\normalfont (Subsystem Lipschitz conditions).  
The static I/O characteristic $h_i(r_i,w_i;\varepsilon_i)$ is Lipschitz continuous in $r_i \in \bar{\mathcal{R}}_i$ uniformly in $ (w_i,\varepsilon_i) \in \mathcal{W}_i \times (0,\varepsilon_i^*]$. The disturbance I/O gain function $\psi_i(r_i^+,w_i^+,r_i^-,w_i^-;\varepsilon_i)$ is Lipschitz continuous in $r_i^-,r_i^+ \in \bar{\mathcal{R}}_i$ uniformly in $w_i^-,w_i^+ \in \mathcal{W}_i$ and $\varepsilon_i \in (0,\epsilon_i^*]$. In addition, $\psi_i^*$ is sub-linear in that there exists a non-negative function $a_i(r_i)$ such that $|\psi_i^*(w_i^+,w_i^-;r_i^*,\varepsilon_i)-\psi_i^*(0,0;r_i^*,\varepsilon_i)| \leq a_i(r_i^*)|w_i^\pm|$ uniformly in $\bar{\mathcal{R}}_i \times (0,\epsilon_i^*]$.
\hfill $\triangledown$
\end{assumption}

\subsection{NDD for networks composed of monotone subsystems}
With reference to Fig.\ref{fig:Intro}, the perturbed network $\mathcal{N}$ can be regarded as a feedback interconnection of $\mathcal{N}_0$ and $\Delta$. 
The nominal network $\mathcal{N}_0$, with input $w$ and output $d$, has the convergent-input-convergent-output property. Specifically, its I/O gain function can be approximated by $\psi^*:=[\psi^*_1,\cdots,\psi^*_N]^\top$, 
which is composed of subsystem I/O gain functions, as the next Lemma shows.

\begin{lemma} \label{Lemma:ApproximateIOG}
\normalfont Consider $\mathcal{N}_0$ under Assumptions \ref{Ass:Setup.SubsysStability}-\ref{Ass:monotone.SubsysCooperative},\ref{Ass:IntendedInteractions},\ref{Ass:LipschitzConditions}, and suppose that the nominal reference input $r^*$ satisfies $r_i^* \in \text{int}(\bar{\mathcal{R}}_i)$ for all $i$. Then, there exists functions $P,Q:\mathbb{R}_{\geq 0} \times \mathbb{R}_{>0} \to \mathbb{R}_{\geq 0}$, such that if $w(t) \to [w^-,w^+]$, then 
\begin{subequations} \label{Eqn:ApproxIOG}
    \begin{align} 
        d(t) &\to [\psi^*(w^-,w^+;r^*,\varepsilon)-P(|w^\pm|;\varepsilon),\nonumber \\
        &\hspace{40pt}\psi^*(w^+,w^-;r^*,\varepsilon)+P(|w^\pm|;\varepsilon)],\label{Eqn:DistIOG}\\
        y(t) &\to [H(r^*)-Q(|w^\pm|;\varepsilon),H(r^*)+Q(|w^\pm|;\varepsilon)].\label{Eqn:OutputIOG}
    \end{align}
\end{subequations}
Particularly, the functions $P,Q$ can be decomposed as
\begin{equation} \label{Eqn:IOGApproxError}
    \begin{aligned}
    P(|w^\pm|;\varepsilon):&=p_1(\varepsilon)|w^\pm|+p_0(\varepsilon),\\
    Q(|w^\pm|;\varepsilon):&=q_1(\varepsilon)|w^\pm|+q_0(\varepsilon),
\end{aligned}
\end{equation}
where $p_1,p_0,q_1,q_0$ are non-negative scalar functions with the following property:  for each $i$, given any $\mu>0$, there exists $\varepsilon_i^{**}=\varepsilon_i^{**}(\mu,\varepsilon_{i+1},\cdots,\varepsilon_{N})>0$, such that  $p_1(\varepsilon),p_0(\varepsilon),q_1(\varepsilon),q_0(\varepsilon) \leq \mu$ if $0<\varepsilon_i \leq \varepsilon_i^{**}(\mu,\varepsilon_{i+1},\cdots,\varepsilon_{N})$ for all $i$.
\end{lemma}

\noindent The proof of Lemma \ref{Lemma:ApproximateIOG} is in Appendix Section \ref{Sec:ApproximateIO}. 
Essentially, this property holds because each subsystem has the disturbance attenuation property (Assumption \ref{Ass:Setup.DistAttn}) and is monotone (Assumption \ref{Ass:monotone.SubsysCooperative}).
Equation (\ref{Eqn:DistIOG}) allows us to approximate the disturbance I/O behavior of network $\mathcal{N}_0$ using the disturbance I/O gain functions ($\psi_i^*$) of the subsystems $\Sigma_i$ with a constant reference input $r_i^*$. In addition, by (\ref{Eqn:OutputIOG}) and (\ref{Eqn:IOGApproxError}), for a constant $|w^\pm|$, the effect of disturbance input $w$ on the prescribed output $y$ can be arbitrarily diminished by decreasing each $\varepsilon_i$. Yet, for the perturbed network $\mathcal{N}$, we need to prove that $w=w(\varepsilon)$ does not grow as $\varepsilon$ is decreased.
To this end, we need some results on boundedness of discrete time systems. Specifically, consider
\begin{align} \label{Eqn:GeneralDT}
    x(k+1)=F(x(k)),
\end{align}
where $x \in \mathbb{R}^n$, and without loss of generality, we assume that $F(0)=0$. System (\ref{Eqn:GeneralDT}) is said to be ultimately bounded~\cite{HassanKhalilBook} in a box $[x_*^-,x_*^+]$ if, for any initial condition $x(0)$, there exists a $k_*>0$ such that $x(k) \in [x_*^-,x_*^+]$ for all $k \geq k_*$. We use $x(k) \to [x_*^-,x_*^+]$ to denote that $x(k)$ is ultimately bounded in $[x_*^-,x_*^+]$. We next introduce a Lyapunov characterization of the ultimate boundedness property that is robust to perturbations.

\begin{definition} \label{Def:ExpGUB}
\normalfont System (\ref{Eqn:GeneralDT}) is said to be \textit{exponentially ultimately bounded} if there exist positive constants $c_1,c_2,c_3,r_0$ and a function $V(\cdot):\mathbb{R}^n \to \mathbb{R}$ such that
\begin{subequations} \label{Eqn:NominalLyapunov}
    \begin{align}
    c_1 |x|^{2} \leq V(x) &\leq c_2 |x|^{2},\label{Eqn:NominalLyapunov1}\\
    |V(x_1)-V(x_2)| &\leq c_3 |x_1-x_2|\cdot(|x_1|+|x_2|),\\
    V(F(x))- V(x) &\leq -c_4 |x|^{2}, \;\;\text{for all } |x|
    \geq r_0.\label{Eqn:NominalLyapunov2}
    \end{align}
\end{subequations}
Specifically, if (\ref{Eqn:NominalLyapunov}) is satisfied, system (\ref{Eqn:GeneralDT}) is exponentially ultimately bounded in $[-r_*,r_*]$, where $r_*:=c_1 r_0/c_2$. 
\hfill $\triangledown$
\end{definition}
\vspace{-2pt}
\noindent 
If (\ref{Eqn:NominalLyapunov}) is satisfied with $r_0=0$, system (\ref{Eqn:GeneralDT}) has an exponentially stable equilibrium point at $x=0$. 
The boundedness property of an exponentially ultimately bounded system is robust to perturbations. In fact, consider a perturbation of the nominal system (\ref{Eqn:GeneralDT}):
\begin{align} \label{Eqn:GeneralDTPert}
    x(k+1)=F(x(k))+p\cdot \delta(x(k)) ,
\end{align}
where $p$ is a constant parameter and $|\delta(x)| \leq L_1|x|+L_2$ for all $x$. Assume that $F(x)$ is sub-linear, that is, there exists $L_F > 0$ such that $|F(x)| \leq L_F|x|$, then we can prove the following robust boundedness result for the perturbed discrete time system (\ref{Eqn:GeneralDTPert}). 
\begin{lemma}
\label{Lemma:RobustnessGUB}
\normalfont Suppose the nominal system (\ref{Eqn:GeneralDT}) is exponentially ultimately bounded in $[-r_*,r_*]$, then there exists $p_*,\kappa>0$, such that for all $p \in [0,p_*]$, system (\ref{Eqn:GeneralDTPert}) satisfies $x(k) \to [-r_*-\kappa p,r_*+\kappa p]$.
\hfill $\triangledown$
\end{lemma}
\noindent The proof of Lemma \ref{Lemma:RobustnessGUB} can be found in Appendix Section \ref{Sec:Appdx.GUB}. Now we are ready to state our first main result. It uses the monotonicity properties of $\Delta$ and the convergent-input-convergent-output of $\mathcal{N}_0$ in Lemma \ref{Lemma:ApproximateIOG} to provide an $\varepsilon$-independent bound on $w(t;\varepsilon)$, which allows each subsystem to decrease $\varepsilon_i$ for disturbance attenuation. 
\vspace{-2pt}
\begin{theorem} \label{Thm:Monotone}
\normalfont Consider the perturbed network (\ref{Eqn:Setup.Subsys}), (\ref{Eqn:Setup.Intended}), and (\ref{Eqn:Setup.unintended}) under Assumptions
\ref{Ass:Setup.SubsysStability}-\ref{Ass:LipschitzConditions}. Suppose that there exists a set $\mathcal{R}_{\mathcal{N}} \subseteq \prod_{i=1}^N \bar{\mathcal{R}}_i$ and a positive constant vector $\bar{\varepsilon}_0 \leq \varepsilon^*:=[\varepsilon^*_1,\cdots,\varepsilon^*_N]^\top$ such that for each fixed $0<\varepsilon \leq \varepsilon_0$,
$w(t;\varepsilon)$ is bounded for all $t$ and that the discrete time dynamical system
\begin{equation}\label{Eqn:monotone.DTSys}
    \begin{aligned} 
    w^-(k+1) &= \Delta \circ \psi^*(w^-(k),w^+(k);r^*,\varepsilon),\\
    w^+(k+1) &= \Delta \circ \psi^*(w^+(k),w^-(k);r^*,\varepsilon)
    \end{aligned}
\end{equation}
is exponentially ultimately bounded in an $\varepsilon$-independent set $[w_*^-(r^*),w_*^+(r^*)]$ for all $0<\varepsilon \leq \varepsilon_0$ and for every $r^* \in \mathcal{R}_{\mathcal{N}}$. Then, there exists a positive function $\varepsilon^{**}_i(\mu,\varepsilon_{i+1},\cdots,\varepsilon_N)$, such that for any $\mu>0$, $\mathcal{N}$ has the $\mu$-NDD property if $r^* \in \text{int}(\mathcal{R}_{\mathcal{N}})$ and if $0<\varepsilon_i\leq \varepsilon^{**}_i$ for all $i$.
\hfill $\triangledown$
\end{theorem}

\begin{proof}
By Lemma \ref{Lemma:CICS}, $\mathcal{N}_0$ has the convergent-input-convergent-output property.
Since $\Delta$ is cooperative (Assumption \ref{Ass:UnintendedInteractions}) and $w(t)$ is bounded for all $t$, a small-gain theorem for convergent-input-convergent-output systems (Appendix Section \ref{Sec:SGT}) shows that $w(t) \to [w_{**}^-,w_{**}^+]$ if the discrete time system
\begin{equation} \label{Eqn:DTSmallGain}
    \begin{aligned} 
    w^+(k+1)&=\Delta \circ [\psi^*(w^+,w^-;r^*,\varepsilon)+P(|w^\pm(k)|;\varepsilon)],\\
    w^-(k+1)&=\Delta \circ [\psi^*(w^-,w^+;r^*,\varepsilon)-P(|w^\pm(k)|;\varepsilon)],
\end{aligned}
\end{equation}
is ultimately bounded in $[w_{**}^-,w_{**}^+]$. 
To show that $w_{**}^-$ and $w_{**}^+$ can be chosen independent of $\varepsilon$, we treat (\ref{Eqn:DTSmallGain}) as a perturbation of the nominal discrete time system (\ref{Eqn:monotone.DTSys}). By Lemma \ref{Lemma:RobustnessGUB} and with reference to (\ref{Eqn:IOGApproxError}), there exists a $p^*>0$, such that if (\ref{Eqn:monotone.DTSys}) is exponentially ultimately bounded in an $\varepsilon$-independent set $[w_*^-,w_*^+]$ and $|p_1(\varepsilon)|,|p_0(\varepsilon)| \leq p^*$, then $[w_{**}^-,w_{**}^+]$ is $\varepsilon$-independent. Therefore, if $0<\varepsilon_i \leq \min\lbrace\bar{\varepsilon}_0,\varepsilon_i^{**}(p^*,\varepsilon_{i+1},\cdots,\varepsilon_N)\rbrace$ for every $i$,
we can apply Lemma \ref{Lemma:ApproximateIOG} to find
\begin{align*}
    y(t;\varepsilon,\Delta) &\to [H(r^*)-q_1(\varepsilon)|w^\pm_{**}|-q_0(\varepsilon),\\
    & \hspace{32pt}H(r^*)+q_1(\varepsilon)|w^\pm_{**}|+q_0(\varepsilon)],\\
    y(t;\varepsilon,0) &\to [H(r^*)-q_0(\varepsilon),H(r^*)+q_0(\varepsilon)].
\end{align*}
Hence, $\limsup_{t \to \infty}|y(t;\varepsilon,0)-y(t;\varepsilon,\Delta)| \leq q_1(\varepsilon)|w_{**}^\pm|+2q_0(\varepsilon)$, where $w_{**}^\pm$ is $\varepsilon$-independent. This implies that, given any $\mu>0$, $\mu$-NDD can be achieved if each $\varepsilon_i$ is taken sufficiently small such that $q_1(\varepsilon)|w_{**}^\pm|+2q_0(\varepsilon) \leq \mu$.
\end{proof}

Under the conditions of Theorem \ref{Thm:Monotone}, NDD of the $(nN)$-dimensional continuous time system $\mathcal{N}$ can be certified if the $(2N)$-dimensional discrete time system (\ref{Eqn:monotone.DTSys}) is ultimately bounded in an $\varepsilon$-independent set. 
This discrete time system can be constructed using the static disturbance I/O gain functions of the constituent subsystems and the unintended interaction $\Delta$. 
It provides an upper bound for the ``steady state amplification'' of disturbance signals in the perturbed network.
If the trajectory of (\ref{Eqn:monotone.DTSys}) is ultimately bounded in an $\varepsilon$-independent set, then NDD can be achieved if each $\varepsilon_i$ is sufficiently small. 
We call $\mathcal{R}_{\mathcal{N}}$ the \textit{network admissible reference input set} because 
if the subsystems and the prescribed interactions are designed such that $r^* \in \mathcal{R}_{\mathcal{N}}$, then $\mu$-NDD can be achieved for arbitrarily small $\mu$ by decreasing $\varepsilon_i$.
\begin{remark}
\normalfont 
The discrete time dynamical system (\ref{Eqn:monotone.DTSys}) does not explicitly involve the prescribed interaction map $G$. Instead, one can first compute $r^*$ 
assuming no unintended interactions, and then substitute $r^*$ into (\ref{Eqn:monotone.DTSys}) to check if it leads to an ultimately bounded, $\varepsilon$-independent $w(k)$. 
\hfill $\triangledown$
\end{remark}
\begin{remark}
\normalfont Note that $\varepsilon^*_i$ is a function of $\varepsilon_{i+1},\cdots,\varepsilon_N$. This implies that the requirement on disturbance attenuation for an upstream subsystem $i$ is generally stricter than its downstream subsystems $j \geq i+1$ to diminish propagation of the regulation error via prescribed interactions. In the special case where $G(y)\equiv r^*$ (i.e., no prescribed interactions), $\varepsilon_i^*$ can be chosen independent of $\varepsilon_{i+1},\cdots,\varepsilon_N$.
\hfill $\triangledown$
\end{remark}
\begin{remark}
\normalfont The requirement for $w(t;\varepsilon)$ to be bounded for all $t$ for each fixed $\varepsilon$ is often satisfied in physical systems with nonlinear dynamics. 
For example, in biomolecular systems, the state variables represent molecular concentrations, which are often bounded above by conservation laws. 
If an $\varepsilon$-independent bound for $w(t;\varepsilon)$ can be easily found for $\mathcal{N}$, then there is no need to check the boundedness of (\ref{Eqn:monotone.DTSys}). 
\hfill $\triangledown$
\end{remark}

While many engineering subsystems have I/S monotone dynamics, the presence of controllers is often required for them to achieve asymptotic static disturbance attenuation. When a dynamic negative feedback controller is used to regulate a subsystem, the resultant dynamics of the regulated subsystem is often not monotone. 
\begin{example} \label{Example:Non-Monotone}
\normalfont Suppose that a  plant has I/S monotone dynamics $\dot{x}_i=-x_i+u_i+w_i$, $y_i=x_i$, where $u_i=z_i$ is a control input arising from 
a dynamic feedback controller $\dot{z}_i=-z_i+(r_i-x_i)/\varepsilon_i$. It is easy to show that the regulated subsystem
\begin{align} \label{Eqn:RegulatedSubsysExample}
    \dot{x}_i=-x_i+z_i+w_i,\;\dot{z}_i=-z_i+(r_i-x_i)/\varepsilon_i
\end{align}
has the asymptotic static disturbance attenuation property with a nominal static I/O characteristic $H_i(r_i)=r_i$. However, 
the incidence graph induced by
\begin{align*}
    f_i(x_i,z_i,r_i,w_i;\varepsilon_i)=
    \begin{bmatrix}
    -x_i+z_i+w_i\\
    -z_i+(r_i-x_i)/\varepsilon_i
    \end{bmatrix}
\end{align*} 
contains a negative cycle $x_i \inhib z_i \to x_i$, indicating that it is non-monotone according to Lemma \ref{Lemma:graphical}.
\hfill $\triangledown$
\end{example} 
\noindent Motivated by this example, in the next section, we seek conditions for networks composed of non-monotone subsystems to achieve NDD. In the context of Example \ref{Example:Non-Monotone}, we show that if the dynamics of the feedback controller $z_i$ is sufficiently fast, then (\ref{Eqn:RegulatedSubsysExample}) behaves like an I/S monotone system, thus, the results developed in this section hold with similar conditions.

\section{Network disturbance decoupling with two-timescale non-monotone subsystems} \label{Sec:SP-monotone}
Certain non-monotone systems can have dynamic properties similar to those of monotone systems~\cite{Wang2008,Sootla2018}. In particular, for autonomous systems, if the ``non-monotone dynamics'' in a two-timescale non-monotone system evolve at a sufficiently fast rate, certain convergence properties for monotone systems are preserved~\cite{Wang2008}. Based on similar reasonings, we provide conditions for NDD of networks composed of non-monotone subsystems.

\subsection{Two-timescale subsystem setup}
\noindent We consider subsystem $\Sigma_i$ parameterized by an additional small positive parameter $\nu$, which induces a timescale separation in the subsystems. For simplicity, we use the same $\nu$ for all subsystems, although the results are not restricted to this case. We now write $\Sigma_i=\Sigma_i(\varepsilon_i,\nu)$ as:
\begin{equation}\label{Eqn:RegulatedSubsystem}
    \begin{aligned} 
    \dot{x}_i&=f_i(x_i,z_i,u_i;\varepsilon_i),&
    y_i&=l_i(x_i)&\\
    \nu \dot{z}_i&=g_i(x_i,z_i,u_i;\varepsilon_i),&
    d_i&=\rho_i(x_i,z_i),& 
    \end{aligned}
\end{equation}
where $x_i \in \mathcal{X}_i \subseteq \mathbb{R}^n$, $z_i \in \mathcal{Z}_i \subseteq \mathbb{R}^m$ and the I/O signals $u_i=[r_i,w_i]^\top$ and $q_i=[y_i,d_i]^\top$ are defined as before in Section \ref{Sec:ProblemFormulation}. We assume that the prescribed output $y_i$ is a function of the slow state $x_i$ only, but the disturbance output $d_i$ may depend on both $x_i$ and $z_i$.
Subsystem (\ref{Eqn:RegulatedSubsystem}) is singularly perturbed by $\nu$. In particular, in the fast time scale $\tau=t/\nu$, the boundary layer dynamics~\cite{HassanKhalilBook} of (\ref{Eqn:RegulatedSubsystem}) are:
\begin{align} \label{Eqn:SubsystemBL}
    \mathrm{d}z_i/\mathrm{d}\tau=g_i(x_i,z_i,u_i;\varepsilon_i),
\end{align}
where $x_i$ and $u_i$ are treated as fixed parameters.
\begin{assumption} \label{Ass:BLeq}
\normalfont (Subsystem boundary layer). For every fixed $(x_i,u_i) \in \mathcal{X}_i \times (\mathcal{R}_i \times \mathcal{W}_i)$ and $\varepsilon_i \in (0,\varepsilon_i^*]$, system (\ref{Eqn:SubsystemBL}) has a GAS equilibrium $\bar{z}_i=\Gamma_i(x_i,u_i;\varepsilon_i) \in \mathcal{Z}_i$.
\hfill $\triangledown$
\end{assumption}
\noindent Substituting $z_i=\Gamma_i(x_i,u_i;\varepsilon_i)$ into (\ref{Eqn:RegulatedSubsystem}), a candidate reduced model of (\ref{Eqn:RegulatedSubsystem}) is:
\begin{align}\label{Eqn:SubsystemReduced}
    \dot{\bar{x}}_i=\bar{f}_i(\bar{x}_i,u_i;\varepsilon_i),\;\;
    \bar{d}_i = \bar{\rho}_i(\bar{x}_i,u_i;\varepsilon_i),\;\;
    \bar{y}_i = l_i(\bar{x}_i),
\end{align}
where $\bar{f}_i(\bar{x}_i,u_i;\varepsilon_i):=f_i(\bar{x}_i,\Gamma_i(\bar{x}_i,u_i;\varepsilon_i),u_i;\varepsilon_i)$ and $\bar{\rho}_i(\bar{x}_i,u_i;\varepsilon_i):=\rho_i(\bar{x}_i,\Gamma_i(\bar{x}_i,u_i;\varepsilon_i))$.
We denote system (\ref{Eqn:SubsystemReduced}) by $\bar{\Sigma}_i$ and require it to have similar stability, disturbance attenuation, monotonicity, and Lipschitz continuity properties as specified for the subsystems in Section \ref{Sec:MonotoneSubsystems}. These conditions are summarized below.
\begin{assumption} \label{Ass:SP.Reduced}
\normalfont Each $\bar{\Sigma}_i$ satisfies the following: 
\begin{enumerate}[label=(\roman*)]
    \item It is I/S monotone with respect to the partial orders $(\sigma^u;\sigma^x)$  for all $\varepsilon_i \in (0,\varepsilon_i^*]$. The output functions $\bar{\rho}_i$ and $l_i$ have sign-stable partial derivatives.
    \item It is endowed with a well-defined I/S characteristic $\bar{\varphi}_i(u_i;\varepsilon_i)$. The I/O characteristics $h_i(u_i;\varepsilon_i)$ satisfies Assumption \ref{Ass:Setup.DistAttn}.
    \hfill $\triangledown$
\end{enumerate}
\end{assumption}
\noindent By this assumption, the functions $\bar{\varphi}_i(u_i;\varepsilon_i)$ and $\bar{\rho}_i(x_i,z_i)$ 
have canonical decomposition functions $\hat{\varphi}_i(u_i^+,u_i^-;\varepsilon_i)$ and $\hat{\rho}_i(\bar{x}_i^{+},u_i^+,\bar{x}_i^{-},u_i^-;\varepsilon_i)$, respectively.
The decomposition functions can be composed according to Lemma \ref{Lemma:ComposeMixedMonotone} to obtain the disturbance I/O gain function
\begin{align*}
    \psi_i(u_i^+,u_i^-;\varepsilon_i)=
    \hat{\rho}_i(\hat{\varphi}_i(u_i^+,u_i^-;\varepsilon_i),u_i^+,\hat{\varphi}_i(u_i^-,u_i^+;\varepsilon_i),u_i^-;\varepsilon_i).
\end{align*}
Similar to (\ref{Eqn:monotone.fixedinputIOG}), we write 
\begin{align} \label{Eqn:SP.fixedinputIOG}
    \psi_i^*(w_i^+,w_i^-;r_i^*,\varepsilon_i):=\psi_i(r_i^*,w_i^+,r_i^*,w_i^-;\varepsilon_i)
\end{align}
for the subsystem static disturbance I/O gain function for a fixed reference input $r_i^*$.
Under mild technical conditions, the conditions to guarantee NDD in Theorem \ref{Thm:Monotone} for networks with monotone subsystems are also sufficient for networks composed of two-timescale subsystems (\ref{Eqn:RegulatedSubsystem}). 
To show this, we extend the convergent-input-convergent-state/output results for monotone systems in Lemma \ref{Lemma:CICS} to singularly perturbed systems with monotone reduced dynamics. 
This requires an additional technical assumption as follows. 

\begin{assumption}\label{Ass:BoundedDerivative}
\normalfont 
There exists $M_1(\varepsilon)>0$, independent of $\nu$, such that $|u(t)| \leq M_1(\varepsilon)$ for all $t$. In addition, there exists $M_2(\varepsilon)>0$, independent of $\nu$, such that $\|\dot{u}\| \leq M_2(\varepsilon)$. 
\hfill $\triangledown$
\end{assumption}
\begin{lemma} \label{Lemma:SPCICS}
\normalfont
(Approximate convergent-input-convergent-output for singularly perturbed monotone systems). 
Consider system (\ref{Eqn:RegulatedSubsystem})
and suppose that Assumptions \ref{Ass:BLeq}-\ref{Ass:BoundedDerivative} are satisfied. 
Then, given any $e>0$, there exists $\nu^*(e;\varepsilon)$, 
such that for a fixed $\varepsilon$, if $0<\nu \leq \nu^*$ and
$u_i(t) \to [u_i^-,u_i^+]$, then
\begin{align} \label{Eqn:CICO}
    d_i(t) \xrightarrow{e} [\psi_i(u_i^-,u_i^+;\varepsilon_i),\psi_i(u_i^+,u_i^-;\varepsilon_i)].
\end{align}
\end{lemma}

\begin{proof}
See Section \ref{Sec:SPCICS} in the Appendix for details. 
\end{proof}
\begin{remark} \label{Remark:NoDerivativeReq}
\normalfont If $g_i$ is not a function of $u_i(t)$, the requirement $\|\dot{u}_i\| \leq M_2$ in Assumption \ref{Ass:BoundedDerivative} can be removed. 
\hfill $\triangledown$
\end{remark}


\subsection{NDD for networks composed of two-timescale subsystems}
Using Lemma \ref{Lemma:SPCICS}, we can determine conditions for NDD of a perturbed network composed of two-timescale non-monotone subsystems. 


\begin{theorem} \label{Thm:SP}
\normalfont Consider the perturbed network (\ref{Eqn:Setup.Intended}), (\ref{Eqn:Setup.unintended}), and (\ref{Eqn:RegulatedSubsystem}) under Assumptions 
\ref{Ass:UnintendedInteractions}-\ref{Ass:BoundedDerivative}.
Suppose that there exists a set $\mathcal{R}_{\mathcal{N}} \subseteq \prod_{i=1}^N \bar{\mathcal{R}}_i$ and a positive constant $\bar{\varepsilon}_0 \leq \varepsilon^*$ such that
for all $0<\varepsilon \leq \varepsilon_0$ the discrete time system (\ref{Eqn:monotone.DTSys}), where $\psi^*$ is the I/O gain function of the reduced system defined in (\ref{Eqn:SP.fixedinputIOG}),
is exponentially ultimately bounded in an $\varepsilon$-independent set $[w_*^-,w_*^+]$. Then, given any $\mu>0$, $\mu$-NDD can be achieved if $r^* \in \text{int}(\mathcal{R}_{\mathcal{N}})$ and if for all $i$
\begin{align} \label{Eqn:Structural.ParameterSetCL}
  0<\varepsilon_i\leq \varepsilon^{***}_i(\mu,\varepsilon_{i+1},\cdots,\varepsilon_{N}),\;0<\nu \leq \nu^{*}(\mu,\varepsilon),
\end{align}
where $\varepsilon_i^{***}$ and $\nu^{**}$ are both positive functions non-increasing with $\mu$.
\hfill $\triangledown$
\end{theorem}

\begin{proof}
(Sketch).
The proof is similar to that of Theorem \ref{Thm:Monotone} but we need to keep track of the 
model reduction error arising from applying Lemma \ref{Lemma:SPCICS} to the subsystems. In particular, for a perturbed network composed of singularly perturbed monotone subsystems, there exists $\nu^{**}(\mu_1;\varepsilon)$ such that for all $0<\nu \leq \nu^{**}$, the convergence result in (\ref{Eqn:ApproxIOG}) can be replaced by
\begin{subequations}
    \begin{align*} 
        d(t) &\xrightarrow{\mu_1} [\psi^*(w^-,w^+;r^*,\varepsilon)-P(|w^\pm|;\varepsilon),\nonumber \\
        &\hspace{40pt}\psi^*(w^+,w^-;r^*,\varepsilon)+P(|w^\pm|;\varepsilon)],\\
        y(t) &\xrightarrow{\mu_1} [H(r^*)-Q(|w^\pm|;\varepsilon),H(r^*)+Q(|w^\pm|;\varepsilon)],
    \end{align*}
\end{subequations}
where $P$ and $Q$ have the same form as those in (\ref{Eqn:IOGApproxError}). If the discrete time system (\ref{Eqn:monotone.DTSys}) converges to $[w_*^-,w_*^+]$, the small-gain theorem for approximate convergent-input-convergent-output systems (Lemma \ref{Lemma:SGT} in Appendix Section \ref{Sec:SGT}) leads to $w(t) \xrightarrow{\alpha(\mu_1)} [w_{**}^-,w_{**}^+]$, where $\alpha(\cdot)$ is a class $\mathcal{K}_0$ function, and $w_{**}^+$ and $w_{**}^-$ are $\varepsilon$-independent. The rest of the proof is similar to that of Theorem \ref{Thm:Monotone}. One can take, for example, $\varepsilon_i^{***}:=\varepsilon_i^{**}(\mu/2,\varepsilon_{i+1},\cdots,\varepsilon_N)$ and  $\nu^{**}=\nu^{*}(\mu/2,\varepsilon)$.
\end{proof}

\noindent In summary, in addition to the conditions of Theorem \ref{Thm:Monotone}, to achieve NDD for networks composed of non-monotone subsystems, Theorem \ref{Thm:SP} requires that the timescale separation in each subsystem is sufficiently large ($\nu$ is sufficiently small). This ensures that the behavior of $\Sigma_i$, which may be non-monotone, are sufficiently close to that of $\bar{\Sigma}_i$, which is monotone. Since $\nu^{**}$ depends on $\varepsilon$, when decreasing $\varepsilon$ to achieve $\mu$-NDD for a fixed $\mu$, it is important to ensure that $\nu \leq \nu^{**}(\mu,\varepsilon)$ remains satisfied. 
\begin{remark} \label{Remark:Roles_eps_mu}
\normalfont While a small $\varepsilon$ ensures that the equilibrium location of $\mathcal{N}$ is close to that of $\mathcal{N}_0$, the value of parameter $\nu$ does not affect the equilibrium location of (\ref{Eqn:RegulatedSubsystem}) and hence that of the perturbed network. The role of a small $\nu$ is to guarantee that $\mathcal{N}$ is dynamically ``well-behaved''. This is a consequence of the approximate convergent-input-convergent-state property for singularly perturbed monotone subsystems in Lemma \ref{Lemma:SPCICS}.
\hfill $\triangledown$
\end{remark}


\section{Motivating example revisited} \label{Sec:Examples}



Here we apply Theorem \ref{Thm:SP} to a network composed of genetic feedback-regulated subsystems described in Section \ref{Sec:MotivatingExample}. The feedback-regulated subsystem dynamics in (\ref{Eqn:sRNA-Dyn}) are not monotone, because the incidence graph induced by the dynamics in (\ref{Eqn:sRNA-Dyn}) contains a negative cycle: $s_i \inhib m_i \to p_i \to s_i$. 
However, if the decay rate constant $\delta_0$ of the RNA species $\mathrm{m}_i$ and $\mathrm{s}_i$ can be made much larger than the decay rate constant $\delta$ of protein $\mathrm{p}_i$, then the subsystem dynamics can be regarded as a two-timescale system~\cite{Grunberg2019}. In particular, the subsystem dynamics can be re-written as:
\begin{subequations} \label{Eqn:sRNA-Dyn-SP}
    \begin{align}
    \nu\dot{m}_i &= \frac{1}{\varepsilon_i}r_i-\frac{1}{\varepsilon_i} \lambda_i m_i s_i - \delta m_i,\\
    \nu\dot{s}_i &= \frac{1}{\varepsilon_i} \beta_i p_i - \frac{1}{\varepsilon_i} \lambda_i m_i s_i - \delta s_i,\\
    \dot{p}_i & = T_i(m_i,w_i)-\delta p_i,\label{Eqn:sRNA-Dyn-SP-3} \\ 
    y_i&=p_i,\;d_i=\rho_i(x_i,z_i)=m_i/\kappa_i. \label{Eqn:sRNA-Dyn-SP-4}
\end{align}
\end{subequations}
System (\ref{Eqn:sRNA-Dyn-SP}) is in the form of (\ref{Eqn:RegulatedSubsystem}), with fast state variables $z_i=[m_i,s_i]^\top$, slow state variable $x_i=p_i$, reference input $r_i$, disturbance input $w_i$, prescribed output $y_i$, and disturbance output $d_i$. 
In practice, the decay rate constant ($\delta_0$) of mRNA and sRNA is often faster than that of protein ($\delta$)~\cite{Levine2007}. To further increase $\delta_0$ to reduce $\nu$, one can (a) engineer the mRNA sequence to recruit additional RNase for its degradation or (b) produce an additional mRNA species $\mathrm{m}_i'$ that can bind and sequester sRNA $\mathrm{s}_i$ to effectively enhance its removal rate~\cite{DelVecchio2017,Grunberg2019}. The parameter $\varepsilon_i$ can be decreased experimentally by increasing the amount of DNA that encodes $\mathrm{m}_i$ and $\mathrm{s}_i$~\cite{Huang2018} and by rational design of the sRNA sequence~\cite{Na2013}. In order to apply Theorem \ref{Thm:SP}, the following section verifies Assumptions \ref{Ass:UnintendedInteractions}-\ref{Ass:BoundedDerivative}. 

\subsection{Verification of Assumptions \ref{Ass:UnintendedInteractions}-\ref{Ass:BoundedDerivative}}
Recall from (\ref{Eqn:GeneComp}) that ribosome competition can be modeled as unintended interaction
\begin{align} \label{Eqn:sRNA.Unintended}
    w_i=\Delta_i(d)=\sum_{j \neq i} d_j,
\end{align}
which satisfies Assumption \ref{Ass:UnintendedInteractions}. 
The prescribed interactions $G_i$ are Hill functions, which are  globally Lipschitz. 
We only consider $G$ that does not contain any feedback loops and, thus, satisfies Assumption \ref{Ass:IntendedInteractions}. 
These interaction maps and subsystem dynamics (\ref{Eqn:sRNA-Dyn-SP}) give rise to the perturbed gene network $\mathcal{N}$. 
The non-negative orthant is positvely invariant under the dynamics of $\mathcal{N}$.
Hence, we have $\mathcal{X}_i,\mathcal{R}_i,\mathcal{W}_i=\mathbb{R}_{\geq 0}$ and $\mathcal{Z}_i=\mathbb{R}^2_{\geq 0}$.
The required Lipschitz conditions in Assumption \ref{Ass:LipschitzConditions} are verified in Appendix Section \ref{Sec:GlobalLipschitz}.

To verify Assumption \ref{Ass:BLeq}, the boundary layer dynamics are:
\begin{equation}
    \begin{aligned} \label{Eqn:sRNA.Controller}
   \frac{\mathrm{d}}{\mathrm{d}\tau}m_i &= \frac{1}{\varepsilon_i}( r_i-\lambda_i m_is_i)-\delta m_i,&\\
   \frac{\mathrm{d}}{\mathrm{d}\tau}s_i &= \frac{1}{\varepsilon_i}(\beta_i p_i-\lambda_i m_is_i)-\delta s_i.&
\end{aligned}
\end{equation}
For each fixed pair $(r_i,p_i) \in \mathcal{R}_i \times \mathcal{X}_i$ and positive $\varepsilon_i$, system (\ref{Eqn:sRNA.Controller}) has a unique non-negative equilibrium $\bar{z}_i = [\bar{m}_i,\bar{s}_i]^\top \in \mathcal{Z}_i$, where
\begin{align} \label{Eqn:sRNA.ControllerDoseResponse}
    \bar{m}_i(p_i,r_i;\varepsilon_i)=\frac{A_i+\sqrt{A_i^2+4\varepsilon_i^2 \delta^2 \lambda_i r_i}}{2 \varepsilon_i \delta \lambda_i},
\end{align}
with $A_i(p_i,r_i):=r_i\lambda_i-\beta_i\lambda_i p_i-\delta^2\varepsilon_i^2.$
GAS of $\bar{z}_i$ has been shown using a Lyapunov function~\cite{Blanchini2011,Grunberg2019}. 

To verify Assumption \ref{Ass:SP.Reduced}, we substitute $\bar{m}_i$ into (\ref{Eqn:sRNA-Dyn-SP-3}) and (\ref{Eqn:sRNA-Dyn-SP-4}), the reduced subsystem dynamics $\bar{\Sigma}_i$ follow:
\begin{align} \label{Eqn:sRNA-Reduced}
    \dot{\bar{p}}_i=\bar{f}_i(\bar{p}_i,r_i,w_i;\varepsilon_i),\;\;
    \bar{y}_i = \bar{p}_i,\;\;
    \bar{d}_i= \bar{\rho}_i(\bar{p}_i,r_i;\varepsilon_i),
\end{align}
where
\begin{subequations}
    \begin{align}
    \bar{f}_i(\bar{p}_i,r_i,w_i;\varepsilon_i)&= T_i(\bar{m}_i(\bar{p}_i,r_i;\varepsilon_i),w_i)-\delta \bar{p}_i,\label{Eqn:sRNA-fdefine}\\
    \bar{\rho}_i(\bar{p}_i,r_i;\varepsilon_i)&=\bar{m}_i(\bar{p}_i,r_i;\varepsilon_i)/\kappa_i.
\end{align}
\end{subequations}
According to (\ref{Eqn:GeneComp}) and (\ref{Eqn:sRNA.ControllerDoseResponse}), we have: 
    \begin{align} 
    &\frac{\partial T_i}{\partial \bar{m}_i}>0,&
    &\frac{\partial T_i}{\partial w_i}<0,&
    &\frac{\partial \bar{m}_i}{\partial r_i}>0,&
    &\frac{\partial \bar{m}_i}{\partial \bar{p}_i}<0,&. \label{Eqn:sRNA-OutputPartials-pi}
\end{align}
Hence, $\frac{\partial \bar{f}_i}{\partial r_i}=\frac{\partial \bar{f}_i}{\partial T_i}\cdot \frac{\partial T_i}{\partial \bar{m}_i} \cdot \frac{\partial \bar{m}_i}{\partial r_i}>0$ and $\frac{\partial \bar{f}_i}{\partial w_i}=\frac{\partial \bar{f}_i}{\partial T_i}\cdot \frac{\partial T_i}{\partial w_i}<0$. Consequently,  $\bar{\Sigma}_i$ is I/S monotone with respect to the partial orders $(\sigma^r,\sigma^w;\sigma^x)=(1,-1;1)$. Since the output functions have sign-stable partial derivatives, Assumption \ref{Ass:SP.Reduced}-(i) is satisfied. 
To verify Assumption \ref{Ass:SP.Reduced}-(ii), we first show that the scalar reduced dynamics (\ref{Eqn:sRNA-Reduced}) has a well-defined I/S characteristic. For each fixed $(r_i,w_i) \in \mathcal{R}_i \times \mathcal{W}_i$ and $\varepsilon_i>0$, the function $\bar{f}_i(\bar{p}_i;r_i,w_i,\varepsilon_i)$ is monotonically decreasing in $\bar{p}_i$. 
In addition, since 
$\bar{f}_i(0,r_i,w_i;\varepsilon_i) \geq 0$ and $\lim_{\bar{p}_i \to +\infty}\bar{f}_i(\bar{p}_i,r_i,w_i;\varepsilon_i)=-\infty$, the scalar reduced system $\dot{\bar{p}}_i = \bar{f}_i(\bar{p}_i,r_i,w_i;\varepsilon_i)$ has a GAS equilibrium. Let 
\begin{align} \label{Eqn:sRNA-SIOC}
    \bar{p}_i=\bar{\varphi}_i(r_i,w_i;\varepsilon_i)
\end{align}
be the static I/S characteristic of $\bar{\Sigma}_i$, 
since $\bar{y}_i=\bar{p}_i$, the subsystem static I/O characteristic is $h_i = \bar{\varphi}_i$. To verify the static disturbance attenuation property in Assumption \ref{Ass:SP.Reduced}-(ii), we show in Appendix Section \ref{Sec:sRNA-DistAttn} that there exists constants $K_i^1, K_i^2 >0$ such that
\begin{align} \label{Eqn:sRNA.DistAttn}
    |h_i(r_i,w_i;\varepsilon_i)-r_i/\beta_i| \leq \varepsilon_i (K^1_i |w_i| + K^2_i)
\end{align}
for all $(r_i,w_i) \in \bar{\mathcal{R}}_i \times \mathcal{W}_i$ and for $\varepsilon_i$ sufficiently small, where $\bar{\mathcal{R}}_i$ can be taken as any $\varepsilon$-independent compact subset of $(0,\alpha_i \beta_i/\delta)$. 
Hence, comparing (\ref{Eqn:sRNA.DistAttn}) with equation (\ref{Eqn:Setup.DistAttn}), the nominal static I/O characteristic is $H_i(r_i) = r_i/\beta_i$, and $\bar{\mathcal{R}}_i$ is an admissible reference input set. 

Finally, we verify Assumption \ref{Ass:BoundedDerivative}, which requires $r(t)$ and $w(t)$ and their derivatives to be bounded. 
By (\ref{Eqn:sRNA-Dyn-SP-3}) and the comparison lemma, for any initial condition, $p_i(t)$ is globally attracted to the set $[0,\alpha_i/\delta]$. 
With reference to (\ref{Eqn:sRNA-Dyn-SP-3}), because $T_i(m_i,w_i)$ is bounded in $[0,\alpha_i]$, $\dot{p}_i$ is bounded in $[-\alpha_i,\alpha_i]$.
Because $r_i=G_i(y)=G_i(p)$ and $G(\cdot)$ is a Hill function, $\dot{r}_i(t)$ and $r_i(t)$ are both bounded.
Similarly, it is possible to verify from (\ref{Eqn:sRNA-Dyn-SP}) that  $[0,r_i/\varepsilon_i]$ is a globally attractive set for $m_i(t)$ and hence $w_i(t)=\sum_{j \neq i}d_j=\sum_{j \neq i}m_j/\kappa_j$ is bounded by a $\nu$-independent constant. We do not need $\dot{w}(t)$ to be bounded by an $\nu$-independent constant because the boundary layer dynamics (\ref{Eqn:sRNA.Controller}) does not depend on $w(t)$.

\subsection{Application of Theorem \ref{Thm:SP}}
Because $\bar{\Sigma}_i$ is I/S monotone with respect to $(\sigma^r,\sigma^w;\sigma^x)=(1,-1;1)$,
the canonical decomposition function of $\bar{\varphi}_i$, which we denote as $\hat{\varphi}_i$, is
\begin{align} \label{Eqn:sRNA.CanonicalDecompH}
    \hat{\varphi}_i(r_i^+,w_i^+,r_i^-,w_i^-;\varepsilon_i)=\bar{\varphi}_i(r_i^+,w_i^-;\varepsilon_i).
\end{align}
For the disturbance output function $\bar{\rho}_i$, since $\partial \bar{\rho}_i/\partial \bar{p}_i<0, \partial \bar{\rho}_i/\partial r_i>0$, the canonical decomposition function $\hat{\rho}_i$ of $\bar{\rho}_i$ is:
\begin{align} \label{Eqn:sRNA.CanonicalDecompRho}
    \hat{\rho}_i(p_i^+,r_i^+,p_i^-,r_i^-;\varepsilon_i)=\bar{\rho}_i(p_i^-,r_i^+;\varepsilon_i).
\end{align}
Given (\ref{Eqn:sRNA.CanonicalDecompH})-(\ref{Eqn:sRNA.CanonicalDecompRho}), and according to Lemma \ref{Lemma:ComposeMixedMonotone}, a static disturbance I/O gain function of $\bar{\Sigma}_i$ is $\psi_i(r_i^+,w_i^+,r_i^-,w_i^-;\varepsilon_i):=\bar{\rho}_i(\bar{\varphi}_i(r_i^-,w_i^+;\varepsilon_i),r_i^+;\varepsilon_i)=\bar{m}_i(\bar{\varphi}_i(r_i^-,w_i^+;\varepsilon_i),r_i^+;\varepsilon_i)/\kappa_i$. Because of this and by equation (\ref{Eqn:SP.fixedinputIOG}), for a fixed input $r_i^*$, we have:
\begin{align} \label{Eqn:sRNA.IOG1}
    \psi^*_i(w_i^+,w_i^-;r_i^*,\varepsilon_i)=\bar{m}_i(\bar{\varphi}_i(r_i^*,w_i^+;\varepsilon_i),r_i^*,\varepsilon_i)/\kappa_i.
\end{align}
On the other hand, when (\ref{Eqn:sRNA-Reduced}) reaches steady state, $\bar{m}_i$ necessarily satisfies 
\begin{align*}
    \alpha_i \frac{\bar{m}_i(\bar{\varphi}_i(r_i^*,w_i^+;\varepsilon_i),r_i^*,\varepsilon_i)/\kappa_i}{1+\bar{m}_i(\bar{\varphi}_i(r_i^*,w_i^+;\varepsilon_i),r^*_i;\varepsilon_i)/\kappa_i+w_i^+}=\delta \bar{\varphi}_i(r_i^*,w_i^+;\varepsilon_i).
\end{align*}
Substituting into (\ref{Eqn:sRNA.IOG1}), the disturbance I/O gain function of (\ref{Eqn:sRNA-Reduced}) can be written alternatively as:
\begin{align}\label{Eqn:sRNA.IOG}
    \psi^*_i(w_i^+,w_i^-;r_i^*,\varepsilon_i)=\frac{\delta \bar{\varphi}_i(r_i^*,w_i^+;\varepsilon_i)(1+w_i^+) }{\alpha_i - \delta \bar{\varphi}_i(r_i^*,w_i^+;\varepsilon_i)}.
\end{align}

With Assumptions 
\ref{Ass:UnintendedInteractions}-\ref{Ass:BoundedDerivative}
satisfied, we can apply Theorem \ref{Thm:SP} to determine if NDD can be achieved for genetic circuits composed of subsystems (\ref{Eqn:sRNA-Dyn-SP}). 
Specifically, we find that the discrete time dynamical system (\ref{Eqn:monotone.DTSys}), where $\psi^*$ is given by (\ref{Eqn:sRNA.IOG}), is exponentially ultimately bounded in an $\varepsilon$-independent set for some $r^*$ values, and hence NDD can be guaranteed for some $r^*$ according to Theorem \ref{Thm:SP}.

\begin{proposition}\label{Prop:sRNA}
\normalfont 
Let $\mathcal{R}_{\mathcal{N}}$ be an $\varepsilon$-independent compact subset of
\begin{align} \label{Eqn:sRNA.NetworkSet}
\tilde{\mathcal{R}}_{\mathcal{N}}:=\left\lbrace r_i^* \in \bar{\mathcal{R}}_i: \sum_{j \neq i} \frac{r_j^* \delta}{\alpha_j \beta_j - \delta r_j^*} < 1,\forall i \right\rbrace.
\end{align}
Given any $\mu>0$, the perturbed network  (\ref{Eqn:Setup.Intended}), (\ref{Eqn:sRNA-Dyn-SP}), and (\ref{Eqn:sRNA.Unintended}) has the $\mu$-NDD property if $r^* \in \mathcal{R}_{\mathcal{N}}$
and $\nu$ and each $\varepsilon_i$ are sufficiently small.
\hfill $\triangledown$
\end{proposition} 

\begin{proof}
With all assumptions in Theorem \ref{Thm:SP} satisfied, we only need to verify that the discrete time system (\ref{Eqn:monotone.DTSys}) is exponentially ultimately bounded in an $\varepsilon$-independent set. Given the form of $\psi^*_i$ in (\ref{Eqn:sRNA.IOG}), we find that the dynamics of $w^+$ and $w^-$ in  (\ref{Eqn:monotone.DTSys}) will be completely decoupled. Hence, it is sufficient to show that the trajectory of the following $N$-dimensional discrete time system is exponentially ulitmately bounded in an $\varepsilon$-independent set:
\begin{align} \label{Eqn:sRNA.DTGUB}
    w_i(k+1) = &\Delta_i \circ \psi^*(w(k);r^*,\varepsilon)\nonumber \\
    =&\sum_{j\neq i} \frac{\delta \bar{\varphi}_j(r_j^*,w_j(k);\varepsilon_j)(1+w_j(k))}{\alpha_j-\delta \bar{\varphi}_j(r_j^*,w_j(k);\varepsilon_j)}.
\end{align}
To this end, we define
\begin{align*}
    \eta_i(r_i,w_i;\varepsilon_i):=\frac{\delta \bar{\varphi}_i(r_i,w_i;\varepsilon_i)}{\alpha_i-\delta \bar{\varphi}_i(r_i,w_i;\varepsilon_i)},\;
    \eta_i^*(r_i):=\frac{\delta r_i }{\alpha_i \beta_i - \delta r_i},
\end{align*}
and note that for all $(r_i,w_i;\varepsilon_i) \in \bar{\mathcal{R}}_i \times \mathcal{W}_i \times (0,\varepsilon_i^*]$, the followings are satisfied: (i) $\eta_i(r_i,w_i;\varepsilon_i) >0$, (ii) $\eta_i(r_i,w_i;\varepsilon_i)<\eta_i(r_i,0;\varepsilon_i)$, and (iii) $\alpha_i - \delta h_i(r_i,w_i;\varepsilon_i)$ is bounded away from 0, and thus by (\ref{Eqn:sRNA.DistAttn}), there exists constant $k_i>0$ such that $|\eta_i(r_i,0;\varepsilon_i)-\eta_i^*(r_i)| \leq k_i \varepsilon_i$. Using these properties, we consider $V(k):=|w(k)|^2$ as a candidate Lyapunov function, which satisfies
\begin{align}
    V(k+1) &= \left|\sum_{j \neq i} \eta_j(r_j^*,w_j(k);\varepsilon_j)(1+w_j(k))\right|^2 \nonumber \\
    &\leq (1+|w(k)|)^2 \left[\sum_{j \neq i}\eta_j^*(r_j^*)+k_j \varepsilon_j\right]^2.
\end{align}
Because $r^* \in \mathcal{R}_\mathcal{N}$ and $\varepsilon \in (0,\varepsilon_i^*]$, there exists an $\varepsilon$-independent constant $0<\vartheta<1$, such that $\sum_{j \neq i}(\eta_j^*(r_j^*)+k_j \varepsilon_j) \leq 1-\vartheta$ for all $i$. Thus, we have $V(k+1)-V(k) \leq (1-\vartheta)(1+|w|)^2-|w|^2 \leq -\vartheta|w|^2+2(1-\vartheta)|w|+(1-\vartheta) \leq -\vartheta |w|^2/2$ if $|w| \geq w_*:=\max\lbrace 1,6(1-\vartheta)/\vartheta \rbrace$, where $w_*$ is $\varepsilon$-independent. This proves that (\ref{Eqn:sRNA.DTGUB}) is exponentially ultimately bounded in $[0,w_*]$.
\end{proof}

\begin{figure}
    \centering
    \includegraphics[scale=0.36]{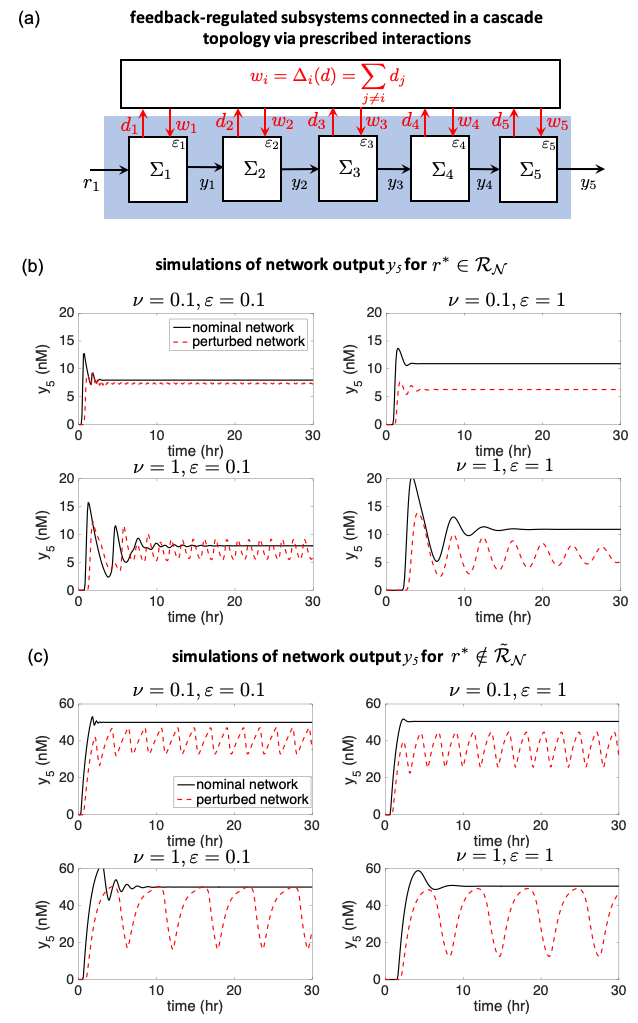}
    \caption{\textbf{Network disturbance decoupling for a cascade of feedback-regulated genetic subsystems.}
    (a) Schematic of a genetic circuit composed of five feedback-regulated subsystems connected in a cascade topology. 
    (b) Simulation results for the network when $r_i^* \in \mathcal{R}_{\mathcal{N}}$. (c) Simulation results for the network when $r_i^* \notin \tilde{\mathcal{R}}_{\mathcal{N}}$. Simulation parameters are identical for all subsystems: $\alpha_i=70$ $\text{nM}/\text{hr}$, $\lambda_i=5$ $(\text{nM}\cdot \text{hr})^{-1}$, $\delta=0.5$ $\text{hr}^{-1}$, $\beta_i=1$ $\text{hr}^{-1}$, $\kappa_i=10$ $\text{nM}$, and $\varepsilon_i=\varepsilon$ for all $i$. The prescribed interactions follow equation (\ref{Eqn:sRNA.RegulatoryInteractions}) with parameters: $n_i=4$, $k_i=6$ $\text{nM}$, and  and $r_1^*=10$ $\text{nM}/\text{hr}$. For panel (b) $B_i=10$ $\text{nM}/\text{hr}$ for all $i \geq 2$ and for panel (c) $B_i=10$ $\text{nM}/\text{hr}$ for $i=2$ and $B_i=50$ $\text{nM}/\text{hr}$ for $i=3,4,5.$ \vspace{-10pt}}  
    \label{fig:Cascade}
\end{figure}

\noindent Therefore, if $G$ and $\Sigma_i$ are designed such that $r^* \in \mathcal{R}_{\mathcal{N}}$, then the network behavior can be made independent of $\Delta$ (i.e., NDD is achieved) by making $\varepsilon_i$ sufficiently small in each subsystem. Our result thus provides an analytical robustness performance limit for genetic circuits composed of feedback-regulated genetic subsystems.

\begin{remark}
\normalfont 
According to (\ref{Eqn:sRNA.NetworkSet}), because $\frac{r_i^*\delta}{\alpha_i\beta_i-\delta r_i^*}$ is positive for $r_i \in \bar{\mathcal{R}}_i$, as the number of subsystems increases, the reference input each subsystem can take for the network to maintain NDD decreases.
\end{remark}
\subsection{Example: Network without prescribed interactions}
\noindent We first consider a network consisting of three identical feedback-regulated subsystems with reference input $r_i^*=r_0$ for all $i$ (see Fig. \ref{fig:IndpInput}a).
Recall that in Fig.\ref{fig:IndpInput}b, our simulations show that NDD can only be achieved for certain reference input levels. 
To explain this, we apply Proposition \ref{Prop:sRNA} and find that any compact subset of $\tilde{\mathcal{R}}_\mathcal{N}:=\lbrace 0<r_0 < 100/3\rbrace$ is a network admissible input set, which we denote by $\mathcal{R}_\mathcal{N}$. 
In accordance with the simulation in  Fig.\ref{fig:IndpInput}b, NDD can be achieved by decreasing $\varepsilon_i$ if $r^* \in \mathcal{R}_{\mathcal{N}}$. On the other hand, decreasing $\varepsilon_i$ does not improve the network's robustness to unintended interactions if $r^* \notin \tilde{\mathcal{R}}_{\mathcal{N}}$, indicating that our result is not conservative. 
The value of $\nu$ does not affect NDD property of $\mathcal{N}$. In fact, local stability of this network can be shown for any $\nu>0$ through linearization~\cite{Qian2018}. Thus, with reference to Remark \ref{Remark:Roles_eps_mu}, there is no need to decrease $\nu$ to ensure network stability and the requirement for $\nu$ to be sufficiently small in Proposition \ref{Prop:sRNA} is conservative in this special case.
\subsection{Example: Network with cascade-like prescribed interactions}
We study another network $\mathcal{N}$ composed of five genetic feedback-regulated subsystems connected in a cascade topology through prescribed interactions, that is, through transcriptional regulation (see Fig. \ref{fig:Cascade}a). 
In particular, we model prescribed interactions as Hill functions~\cite{bfsbook}:
\begin{align} \label{Eqn:sRNA.RegulatoryInteractions}
    r_i=G_i(y_{i-1}) = \begin{cases}
    B_i \frac{(y_{i-1}/k_i)^{n_i}}{1+(y_{i-1}/k_i)^{n_i}},\;&\text{if }i \neq 1,\\
    r^*_1,\;& \text{if }i = 1,
    \end{cases}
\end{align}
where $B_i$ quantifies the maximum transcription rate from gene $i$, $k_i$ is a dissociation constant whose value decreases with the binding affinity between protein $\mathrm{p}_{i-1}$ and the promoter of gene $i$, and $n_i$ is describes the binding cooperativity. 
Using (\ref{Eqn:sRNA.RegulatoryInteractions}) and the subsystem nominal static I/O characteristic $y_i=r_i/\beta_i$, we can compute the nominal reference input $r^*$. 
Simulation results for network $\mathcal{N}$ with different $(\nu,\varepsilon)$ pairs are shown in Fig. \ref{fig:Cascade}b-c.
For the simulations in Fig. \ref{fig:Cascade}b, we choose parameters for the subsystems and the prescribed interaction map such that $r^* \in \mathcal{R}_{\mathcal{N}}$. We therefore apply Proposition \ref{Prop:sRNA} to claim that for arbitrarily small $\mu$, $\mu$-NDD of $\mathcal{N}$ can be achieved by decreasing both $\varepsilon$ and $\nu$, which is consistent with simulations in Fig. \ref{fig:Cascade}b. In contrast, when the parameters are chosen such that $r^* \notin \tilde{\mathcal{R}}_{\mathcal{N}}$, as shown in Fig. \ref{fig:Cascade}c, decreasing $\varepsilon$ and $\nu$ does not lead to NDD.

\section{Discussion and future work}
In this paper, we have studied networked dynamical systems, in which unintended interactions among subsystems perturb the prescribed network's behavior. 
We have provided conditions on subsystem dynamics, the intended and the unintended interaction maps to achieve network disturbance decoupling (NDD), where the steady state outputs from all subsystems become essentially independent of the unintended interactions.
While NDD may be addressed by designing the entire network monolithically, we find that, under certain conditions, NDD can be obtained by simply improving each subsystem's robustness to a constant, state-independent disturbance. 
Specifically, these conditions require that (i) all subsystems are I/S monotone, (ii) the prescribed interactions among subsystems do not contain feedback loops, and (iii) the unintended interactions are cooperative.
When the subsystem dynamics are non-monotone, the same result holds with similar conditions if the subsystem dynamics have a timescale separation property, such that each reduced subsystem dynamics are monotone.
We apply our theoretical result to guide the design of genetic circuits that are robust to context. In particular, we show that a recently implemented biomolecular feedback controller~\cite{Huang2018}, which enables a single genetic subsystem to asymptotically attenuate a constant disturbance, can theoretically be used to regulate multiple genes in a network to reach NDD. 

Experimental validation of the results in Section \ref{Sec:Examples} is underway. In the future, we plan to consider NDD problems for a larger class of unintended interactions $\Delta$, including, for example, $\Delta$ that contain dynamics. 
We also plan to extend this study to multi-stable networks and to consider intended interaction maps that contain feedback loops. 
These studies may provide guidance to engineer networked systems to function robustly in different contexts.

\section{Appendix} \label{Sec:Appendix}
\subsection{Proof of Lemma \ref{Lemma:ApproximateIOG}} \label{Sec:ApproximateIO}
To prove Lemma \ref{Lemma:ApproximateIOG}, we note that the I/O gain function of each subsystem has the following property.

\begin{lemma} \label{Lemma:InequalitiesIOFcn}
\normalfont Suppose that Assumptions \ref{Ass:Setup.SubsysStability},\ref{Ass:Setup.DistAttn},\ref{Ass:LipschitzConditions} are satisfied and let $\hat{h}_i(r_i^+,w_i^+,r_i^-,w_i^-;\varepsilon_i)$ be the canonical decomposition function of $h_i(r_i,w_i;\varepsilon_i)$, then, for any $e_i>0$ such that $r_i-e_i,r_i+e_i \in \bar{\mathcal{R}}_i$,
the function $\hat{h}_i$ satisfies:
\begin{align} \label{Eqn:hLipschitz}
    |\hat{h}_i(r_i+e_i,w_i^+,&r_i-e_i,w_i^-;\varepsilon_i) - H_i(r_i)| \nonumber \\
    &\leq L_h |e_i| + \alpha_i(\varepsilon_i)|w_i^\pm| + \alpha_i^0(\varepsilon_i),
\end{align}
where $L_h>0$ is the Lipschitz constant of $h_i$ .
\hfill $\triangledown$
\end{lemma}

\begin{proof}
Due to Assumption \ref{Ass:monotone.SubsysCooperative}, the static I/O characteristic $y_i=h_i(r_i,w_i;\varepsilon_i)$ is also sign-stable. For $s=r,w$, define $\Lambda_{s}:=\text{sign}(\partial h_i/\partial s_i)$, and let $\Lambda_s^+$ and $\Lambda_s^-$ be defined according to (\ref{Eqn:SignPattern}). 
By equation (\ref{Eqn:DecompFcnConst}), let $\Lambda_{s,j}$ be the $j$-th row of matrix $\Lambda_s$, the canonical decomposition function
\begin{align} \label{Eqn:hDecompFcn}
    \hat{h}_i(r_i^+,w_i^+,r_i^-,w_i^-;\varepsilon_i):=h_i(\mathfrak{p}_r(r_i^\pm),\mathfrak{p}_w(w_i^\pm);\varepsilon_i),
\end{align}
where
\begin{equation}
    \begin{aligned} \label{Eqn:PickFcn}
    \mathfrak{p}_{r,j}(r^\pm) &:= \text{diag}(\Lambda_{r,j}^+)r^+ + \text{diag}(\Lambda_{r,j}^-)r^-,\\
    \mathfrak{p}_{w,j}(w^\pm) &:= \text{diag}(\Lambda_{w,j}^+)w^+ + \text{diag}(\Lambda_{w,j}^-)w^-,\\
\end{aligned}
\end{equation}
are the $j$-th elements of the vector-valued functions $\mathfrak{p}_r$ and $\mathfrak{p}_w$, respectively. Note that (\ref{Eqn:PickFcn}) satisfies $|\mathfrak{p}_{s,j}(s^\pm)| \leq |s^\pm|$. 
Therefore, 
\begin{align} \label{Eqn:wLipschitz}
    |\hat{h}_i(r_i,w_i^+,r_i,w_i^-;\varepsilon_i)&-H_i(r_i)| = |h_i(r_i,\mathfrak{p}_w(w_i^\pm);\varepsilon_i)-H_i(r_i)| \nonumber \\
    & \leq \alpha_i(\varepsilon_i)|\mathfrak{p}_w(w_i^\pm)| + \alpha_i^0(\varepsilon_i)\nonumber \\
    & \leq \alpha_i(\varepsilon_i)|w_i^\pm| + \alpha_i^0(\varepsilon_i).
\end{align}
On the other hand, by the definition of $\hat{h}_i$ in (\ref{Eqn:hDecompFcn}) and the Lipschitz property of $h_i$ in Assumption \ref{Ass:LipschitzConditions}, the decomposition function $\hat{h}_i$ is Lipschitz continuous in $r_i^\pm \in (\bar{\mathcal{R}}_i)^2$ uniformly in $w_i^\pm$ and $\varepsilon_i$ with a Lipschitz constant $L_h$. Hence, we have 
\begin{align} \label{Eqn:rLipschitz}
    |\hat{h}_i(r_i+e_i,w_i^+,r_i-e_i,w_i^-;\varepsilon_i)-&\hat{h}_i(r_i,w_i^+,r_i,w_i^-;\varepsilon_i)| \nonumber \\
    &\leq L_h|e_i|.
\end{align}
Combining (\ref{Eqn:wLipschitz}) and (\ref{Eqn:rLipschitz}), we have (\ref{Eqn:hLipschitz}) proven by triangle inequality.
\end{proof}

\begin{proof}
(Lemma \ref{Lemma:ApproximateIOG}). We prove Lemma \ref{Lemma:ApproximateIOG} through induction. In particular, given $w(t) \to [w^-,w^+]$, we find the ultimate bound for each element of $d(t)$ using the disturbance I/O gain function of each subsystem in (\ref{Eqn:monotone.IOG}), the subsystem static disturbance attenuation property (\ref{Eqn:Setup.DistAttn}), and Assumptions \ref{Ass:IntendedInteractions} and \ref{Ass:LipschitzConditions}. For $i=1$, according to Assumption \ref{Ass:IntendedInteractions}, we necessary have $r_1(t) \equiv r_1^*$, which is independent of the state of all other subsystems. Since $\Sigma_1$ is I/S monotone and the prescribed output function $l_i$ has sign-stable Jacobian, the static I/O characteristic $h_i$ is necessarily equipped with a canonical decomposition function $\hat{h}_i(r_i^+,w_i^+,r_i^-,w_i^-;\varepsilon_i)$ that serves as the I/O gain function for the prescribed output $y$. Thus, if $w_1(t) \to [w_1^-,w_1^+]$, then we have
\begin{equation} \label{Eqn:ApproxIOG1}
    \begin{aligned}
    y_1(t) &\to [\hat{h}_1(r_1^*,w_1^-,r_1^*,w_1^+;\varepsilon_1),\hat{h}_1(r_1^*,w_1^+,r_1^*,w_1^-;\varepsilon_1)],\\
    d_1(t) &\to [\psi_1(r_1^*,w_1^-,r_1^*,w_1^+;\varepsilon_1),\psi_1(r_1^*,w_1^+,r_1^*,w_1^-;\varepsilon_1)].
\end{aligned}
\end{equation}
Let $y_1^*:=H_1(r_1^*)$, by Lemma \ref{Lemma:InequalitiesIOFcn}, we can write
\begin{align} \label{Eqn:ApproxIOG3}
    y_1(t) \to [y_1^*-Q_1(w_1^\pm;\varepsilon_1),y_1^*+Q_1(w_1^\pm;\varepsilon_1)],
\end{align}
where $Q_1(w_1^\pm;\varepsilon_1):=\alpha_1(\varepsilon_1)|w_1^\pm|+\alpha_1^0(\varepsilon_1)$. On the other hand, by the definition of $\psi^*_i$ in (\ref{Eqn:monotone.fixedinputIOG}), the convergence result for $d_1(t)$ in (\ref{Eqn:ApproxIOG1}) can be re-written as $d_1(t) \to [\psi_1^*(w_1^-,w_1^+;r_1^*,\varepsilon_1),\psi_1^*(w_1^+,w_1^-;r_1^*,\varepsilon_1)]$.
Due to Assumption \ref{Ass:IntendedInteractions}, the reference input $r_2=G_2(y)$ to $\Sigma_2$ is only a function of $y_1$. Let $r_2^*:=G_2(y_1^*)$, let $L_G$ be the Lipschitz constant of $G(\cdot)$, we have
\begin{align}
    r_2(t) \to [r_2^*-L_G Q_1(w_1^\pm;\varepsilon_1),r_2^*+L_G Q_1(w_1^\pm;\varepsilon_1)].
\end{align}
We use $r_2^-:=r_2^*- L_G Q_1(w_1^\pm;\varepsilon_1)$ and $r_2^+:=r_2^*+ L_G Q_1(w_1^\pm;\varepsilon_1)$ to denote the ultimate bounds for $r_2(t)$. Since $r_2^*\in\text{int}(\bar{\mathcal{R}}_2)$, for sufficiently small $\varepsilon_1$, $r_2^\pm \in \bar{\mathcal{R}}_2$. 
Similar to our treatment in (\ref{Eqn:ApproxIOG1}) for $\Sigma_1$, we have
\begin{equation}
    \begin{aligned} \label{Eqn:ApproxIOG2}
    y_2(t) &\to [\hat{h}_2(r_2^-,w_2^-,r_2^+,w_2^+;\varepsilon_2),\hat{h}_2(r_2^+,w_2^+,r_2^-,w_2^-;\varepsilon_2)],\\
    d_2(t) &\to [\psi_2(r_2^-,w_2^-,r_2^+,w_2^+;\varepsilon_2),\psi_2(r_2^+,w_2^+,r_2^-,w_2^-;\varepsilon_2)].
\end{aligned}
\end{equation}
By the subsystem disturbance attenuation property (\ref{Eqn:Setup.DistAttn}), let $y_2^*:=H_2(r_2^*)$, we have
\begin{align} \label{Eqn:ApproxIOG4}
    y_2(t) \to [y_2^*-Q_2(w_{\leq 2}^\pm;\varepsilon_{\leq 2}),y_2^*+Q_2(w_{\leq 2}^\pm;\varepsilon_{\leq 2})]
\end{align}
where
\begin{align*}
    Q_2(w_{\leq 2}^\pm;\varepsilon_{\leq 2}):=L_hL_GQ_1(w_1^\pm;\varepsilon_1)+\alpha_2(\varepsilon_2)|w_2^\pm|+\alpha_2^0(\varepsilon_2),
\end{align*}
according to Lemma \ref{Lemma:InequalitiesIOFcn}.
Also due to Assumption \ref{Ass:LipschitzConditions}, the convergence of $d_2(t)$ in (\ref{Eqn:ApproxIOG2}) can be re-written as:
\begin{align*}
    d_2(t) \to [\psi_2^*(w_2^-,w_2^+;r_2^*;\varepsilon_2)-P_2,\psi_2^*(w_2^+,w_2^-;r_2^*;\varepsilon_2)+P_2],
\end{align*}
where
\begin{align*}
    P_2 =P_2(w_1^\pm;\varepsilon_2):&= L_\psi(\varepsilon_2) L_G Q_1(w_1^\pm;\varepsilon_1)\\
    &=L_\psi(\varepsilon_2)[\alpha_1(\varepsilon_1)|w_1^\pm|+\alpha_1^0(\varepsilon_1)], \nonumber
\end{align*}
and $L_{\psi}(\varepsilon)$ is the Lipschitz constant of $\psi_i$ for variables $r_i^-$ and $r_i^+$ as stated in Assumption \ref{Ass:LipschitzConditions}. 
Since we do not assume the Lipschitz property of $\psi_i$ to hold uniformly in $\varepsilon_i$, $L_{\psi}$ is in general dependent on $\varepsilon_i$.
Note that, for a fixed $\varepsilon_2$, since $\alpha_1$ and $\alpha_1^0$ are class $\mathcal{K}$ functions, $P_2$ can be made arbitrarily small if $\varepsilon_1$ is sufficiently small. 
Using (\ref{Eqn:ApproxIOG3}) and (\ref{Eqn:ApproxIOG4}) to determine $r_3^\pm$, we can continue the iteration to find the boxes that bounds $r_3(t)$ and $d_3(t)$. After $k$ iterations, let $w_{\leq k}:=[w_1,\cdots,w_k]^\top$ and $\varepsilon_{\leq k}:=[\varepsilon_1,\cdots,\varepsilon_k]^\top$, we have
    \begin{align*}
    y_k(t) &\to [y_k^*-Q_k,y_k^*+Q_k],\\
    d_k(t)
    &\to [\psi_k^*(w_k^-,w_k^+;r_k^*,\varepsilon_k)-P_k,\psi_k^*(w_k^+,w_k^-;r_k^*,\varepsilon_k)+P_k],
\end{align*}
where $y_k^*=H_k(r_k^*)$ 
\begin{align*}
    Q_k(w_{\leq k}^\pm;\varepsilon_{\leq k}):=&
    \sum_{i=1}^k(L_hL_G)^{k-i} \cdot (\alpha_i(\varepsilon_i)|w_i^\pm|+\alpha_i^0(\varepsilon_i)),\\
    P_k(w_{\leq k}^\pm;\varepsilon_{\leq k}):=&L_\psi(\varepsilon_k)\sum_{i=1}^{k-1} L_h^{k-1-i}L_G^{k-i}\cdot(\alpha_i(\varepsilon_i)|w_i^\pm|+\alpha_i^0(\varepsilon_i)).
\end{align*}
Note $Q(w^\pm;\varepsilon)$ and $P(w^\pm;\varepsilon)$ can be arranged as in (\ref{Eqn:IOGApproxError}). 
Specifically, let 
\begin{align*}
    p_{1,k}(\varepsilon_{\leq k})&:=L_\psi(\varepsilon_k)\sum_{i=1}^{k-1}L_h^{k-1-i}L_G^{k-i} \alpha_i(\varepsilon_i),\\
    p_{0,k}(\varepsilon_{\leq k})&:=L_\psi(\varepsilon_k)\sum_{i=1}^{k-1}L_h^{k-1-i}L_G^{k-i} \alpha_i^0(\varepsilon_i),
\end{align*}
we have $p_j(\varepsilon) = [p_{j,1},\cdots,p_{j,N}]^\top$ for $j=0,1$.
Since $\alpha_i$ and $\alpha_i^0$ are class $\mathcal{K}$ functions, for each $k$, given any $\mu>0$, $p_{1,k}\leq \mu$, and hence $p_1 \leq \mu$, can be satisfied if 
\begin{align*}
    \varepsilon_i \leq \alpha_i^{-1}\left( \frac{\mu L_\psi^{-1}(\varepsilon_k)}{(k-1) L_h^{k-1-i} L_G^{k-1}}\right)=:\varepsilon_{i,k}^{**}(\mu,\varepsilon_k)
\end{align*}
$\forall i \leq k-1$, $\forall k$. We can then take
\begin{align*}
    \varepsilon^{**}_i(\mu,\varepsilon_{\geq i+1}):=\min_{k = i+1,\cdots,N}\varepsilon_{i,k}^{**}(\mu,\varepsilon_k).
\end{align*}
A similar upper bound $\varepsilon^{**}$ can be established for $p_{0},q_{1},q_{0} \leq \mu$ to be satisfied. This completes the proof.
\end{proof}

\subsection{Proof of Lemma \ref{Lemma:RobustnessGUB}} \label{Sec:Appdx.GUB}
\begin{proof}
Consider $V(x)$ in (\ref{Eqn:NominalLyapunov}) as a candidate Lyapunov function for the perturbed system, then we have
\begin{align} 
    &\Delta V:=V(F(x)+p \delta(x))-V(x) \nonumber \\
    = &V(F(x)+p \delta(x))-V(F(x))+V(F(x))-V(x) \nonumber\\
    \leq & c_3 p \delta (x)(|F(x)|+|F(x)+p\delta(x)|) -c_4 |x|^2, \;\forall |x| \geq r_0\nonumber\\
    \leq & (-c_4 + p a(p))|x|^2 + p b(p)|x|+p c(p),\;\forall |x| \geq r_0
    \label{Eqn:GUBProof}
\end{align}
where $a(p) = c_3 (2L_1 L_F+pL_1^2)$, $b(p) = 2c_3L_2(pL_1+L_F)$, and $c(p) = c_3 p L_2^2$. By (\ref{Eqn:GUBProof}), there exists $p_*>0$, such that $\Delta V \leq -c_4 |x|^2/2+p(b(p)|x|+c(p))$ for all $p\in[0,p_*]$. 
For such a fixed $p$, take $r_p:=p \cdot \text{max}\left(2L_2 \sqrt{\frac{2c_3}{c_4}}, 8\frac{b(p_*)}{c_4} \right)$, one can verify that $pb(p)|x|, p c(p) \leq c_4|x|^2/8$ for all $|x| \geq r_p$. 
Hence, $\Delta V \leq -c_4|x|^2/4$ for all $|x| \geq r_0+r_p$. 
By Definition \ref{Def:ExpGUB}, for all $p \in [0,p_*]$, the perturbed system (\ref{Eqn:GeneralDTPert}) is exponentially ultimately bounded in $[-c_1(r_0+r_p)/c_2,c_1(r_0+r_p)/c_2]$.
\end{proof}

\subsection{Proof of Lemma \ref{Lemma:SPCICS}} \label{Sec:SPCICS}
We first show that the reduced system is ISS after a coordinate translation, which allows us to use a singular perturbation result for ISS systems~\cite{Christofides1995} to compute the model reduction error for the fast variable $z_i$. 
This is then used to compute the model reduction error for the slow variable $x_i$. 
Since $\nu$ is the singular perturbation parameter and $\varepsilon_i$ is treated as a constant, we do not explicitly spell out $\varepsilon_i$ in the sequel. We also suppress the subscript $i$ for simplicity in this section. For example, we will write $x$ instead of $x_i$.

Recall $\bar{\varphi}(u)$ is the static I/S characteristic of the reduced system. We let $\tilde{x}:=\bar{x}-\bar{\varphi}(0)$ and write the translated reduced system as:
\begin{align} \label{Eqn:ReducedTranslated}
    \dot{\tilde{x}}=\tilde{f}(\tilde{x},u(t)):=\bar{f}(\tilde{x}+\bar{\varphi}(0),u(t)).
\end{align}
\begin{lemma}
\normalfont Under the assumptions of Lemma \ref{Lemma:SPCICS}, the translated reduced system (\ref{Eqn:ReducedTranslated}) is ISS.
\hfill $\triangledown$
\end{lemma}
\begin{proof}
To show that (\ref{Eqn:ReducedTranslated}) is ISS, we first show that it has the asymptotic gain property (see \cite{Sontag1996}), that is, there exists a class $\mathcal{K}_0$ function $\gamma(\cdot)$ such that $\limsup_{t \to \infty}|\tilde{x}| \leq \gamma(\|u\|)$. According to Theorem 1 in ~\cite{Sontag1996}, this asymptotic gain property, combined with the fact that (\ref{Eqn:ReducedTranslated}) is GAS when $u \equiv 0$, is equivalent to (\ref{Eqn:ReducedTranslated}) being being ISS.
Given Assumption \ref{Ass:SP.Reduced}, 
let $\hat{\varphi}(\cdot,\cdot)$ be the canonical decomposition function of $\bar{\varphi}(\cdot)$ and suppose that $\mathcal{U}:=[\underline{u},\overline{u}]$. 
Let $u^-(\|u\|):=\max(-\mathbf{1}_n\|u\|,\underline{u})$, $u^+(\|u\|):=\min(\mathbf{1}_n\|u\|,\overline{u})$, where $\mathbf{1}_n$ is an $n$-vector with all elements being 1. Therefore, the input $u(t)$ to (\ref{Eqn:ReducedTranslated}) satisfies $u(t) \to [u^-(\|u\|),u^+(\|u\|)]$, and by Lemma \ref{Lemma:CICS}, we have
$ \tilde{x} \to [\tilde{\varphi}^-(\|u\|),\tilde{\varphi}^+(\|u\|)],$
where $\tilde{\varphi}^-(\|u\|),\tilde{\varphi}^+(\|u\|):\mathbb{R} \to \mathbb{R}^n$ are defined as:
\begin{align*}
    \tilde{\varphi}^+(\|u\|)&:=\hat{\varphi}(u^+(\|u\|),u^-(\|u\|))-\bar{\varphi}(0),\\
    \tilde{\varphi}^-(\|u\|)&:=\hat{\varphi}(u^-(\|u\|),u^+(\|u\|))-\bar{\varphi}(0).
\end{align*}
Let $\gamma(\|u\|)=\max_{v \leq \|u\|}\max\lbrace|\tilde{\varphi}^+(v)|,|\tilde{\varphi}^-(v)|\rbrace$. Since $\gamma(0)=0$ and it is non-decreasing, it is an asymptotic gain of (\ref{Eqn:ReducedTranslated}). The GAS property of (\ref{Eqn:ReducedTranslated}) when $u=0$ is a consequence of the existence of the I/S characteristic for all $u \in \mathcal{U}$. 
\end{proof}

\noindent Since the convergent-input-convergent-state property we aim to prove is translation-invariant, we will assume in the sequel that $\bar{\varphi}(0)=0$ and hence the reduced system $\bar{\Sigma}$ is ISS.

\begin{lemma} \label{Lemma:Appendix.SPISS}
\normalfont Under the assumptions of Lemma \ref{Lemma:SPCICS}, given any $\mu>0$, there exists $\nu^*=\nu^*(\mu)$, such that
\begin{align} \label{Eqn:FastvariableBound}
    \limsup_{t \to \infty}|z(t)-\Gamma(x,u(t))| \leq \mu
\end{align}
for all $0<\nu \leq \nu^*$. In addition, the trajectory of (\ref{Eqn:RegulatedSubsystem}) is bounded (by an $\mu$-independent constant) for all $t \geq 0$.
\hfill $\triangledown$
\end{lemma}

\noindent Lemma \ref{Lemma:Appendix.SPISS} is adopted from \cite{Christofides1995}, according to which the boundedness condition for $\|\dot{u}\|$ can be removed if $g$ is independent of $u$. 
To show the convergent-input-convergent-output property in Lemma \ref{Lemma:SPCICS}, let $y_b(t):=z(t)-\Gamma(x(t),u(t))$. The dynamics of $x$ in (\ref{Eqn:RegulatedSubsystem}) can be written as:
\begin{align} \label{Eqn:SPSlowPert}
    \dot{x}=F(x,y_b(t),u(t)):=f(x,\Gamma(x,u)+y_b,u).
\end{align}
We treat (\ref{Eqn:SPSlowPert}) as a perturbation of the reduced system, whose dynamics follow
\begin{align} \label{Eqn:SPSlowNominal}
    \dot{x}=F(x,0,u(t))=f(x,\Gamma(x,u),u).
\end{align}
Let $x(t,y_b(t),u(t))$ be the trajectory of (\ref{Eqn:SPSlowPert}), we aim to show that it is close to $x(t,0,u(t))$, the trajectory of (\ref{Eqn:SPSlowNominal}), as $t \to \infty$ for small $\nu$. 
Given that $u(t) \to [u^-,u^+]$, because both systems are I/S monotone with respect to the input $u(t)$, there exists $u^-_*$ and $u^+_*$, which are two corners of the box set $[u^-,u^+]$, such that for all $t$
\begin{subequations}
    \begin{align}
    x(t,y_b(t),u^-_*) &\leq x(t,y_b(t),u(t)) \leq x(t,y_b(t),u^+_*),\\
    x(t,0,u^-_*) &\leq x(t,0,u(t)) \leq x(t,0,u^+_*).
\end{align}
\end{subequations}
Specifically, $u^-_*=u^-_*(u^-,u^+)$ and $u^+_*=u^+_*(u^-,u^+)$ can be found according to (\ref{Eqn:SignPattern})-(\ref{Eqn:DecompFcnConst}). The trajectories of the nominal system satisfies $ \lim_{t \to \infty}x(t,0,u^-_*) = \hat{\varphi}(u^-,u^+),$ and $\lim_{t \to \infty}x(t,0,u^+_*) = \hat{\varphi}(u^+,u^-)$. 
We now show that $\lim_{\nu \to 0}\limsup_{t \to 0}|x(t,y_b(t),u_*^-)-x(t,0,u_*^-)|=0$. 
To this end, we introduce the following lemma. 
\begin{lemma} \label{Lemma:Appendix.Pert}
\normalfont Consider the nominal system $\dot{x}=F(x,0)$ with a GAS equilibrium $x^*$ and the perturbed system $\dot{x}_p=F(x_p,v(t))$. 
Suppose that $F$ is continuous and locally Lipschitz, and the trajectory of the perturbed system is bounded.
For any $e>0$, there exists $\delta>0$, such that if $\limsup_{t \to \infty}|v(t)|<\delta$, then $\limsup_{t \to \infty}|x_p(t)-x^*| \leq e$. 
\hfill $\triangledown$
\end{lemma}
\noindent This lemma can be derived from Proposition II.4 in \cite{Sontag2003}. 
Since the perturbed system is bounded as a consequence of Lemma \ref{Lemma:Appendix.SPISS}, we can apply Lemma \ref{Lemma:Appendix.Pert}. Because of (\ref{Eqn:FastvariableBound}), we have that for any $\mu>0$, there exists sufficiently small $\nu$ such that $\limsup_{t \to \infty}|x(t,y_b(t),u^-_*)-x(t,0,u^-_*)| = \limsup_{t \to \infty}|x(t,y_b(t),u^-_*)-\hat{\varphi}(u^-,u^+)|\leq \mu$. The same claim can be made for $x(t,y_b(t),u^+_*)$.
This shows that for any given $\mu>0$, $x(t) \xrightarrow{\mu}[\hat{\varphi}(u^-,u^+),\hat{\varphi}(u^+,u^-)]$ for sufficiently small $\nu$. Consequently, the disturbance output satisfies $d(t)  \xrightarrow{\mu} [\psi(u^-,u^+),\psi(u^+,u^-)]$ for sufficiently small $\nu$ because the output function $\rho$ is assumed to be Lipschitz and sign-stable.
\hfill $\blacksquare$

\subsection{Small-gain theorem for (approximate) convergent-input-convergent-output system} \label{Sec:SGT}
We state and prove the small-gain theorem for (approximate) convergent-input-convergent-output (CICO) systems. 
For generality, we consider system (\ref{Eqn:ISSys}) with input $u(t)$ and output $q(t)$. 
This system is interconnected with a cooperative function $u=\Delta(q)$, where $\Delta(\cdot)$ is globally Lipschitz with Lipschitz constant $L_{\Delta}$.

\begin{lemma}\label{Lemma:SGT}
\normalfont 
Suppose that system (\ref{Eqn:ISSys}) has the following approximate CICO property: for any $u^-, u^+$, if $u(t) \to [u^-,u^+]$, then $q(t) \xrightarrow{\mu} [\psi(u^-,u^+),\psi(u^+,u^-)]$, where $\mu>0$ is a parameter. 
Assume that there exists $u_0^+$ and $u_0^-$ such that $u(t) \in [u_0^-,u_0^+]$ for all $t$ in the interconnected system. 
If the discrete time dynamical system 
\begin{equation}\label{Eqn:DTSys}
    \begin{aligned} 
    u^-(k+1) &= \Delta \circ \psi(u^-(k),u^+(k)),\\
    u^+(k+1) &= 
    \Delta \circ \psi(u^+(k),u^-(k)).
    \end{aligned}
\end{equation}
is exponentially ultimately bounded in $[u_*^-,u_*^+]$, then there exists $\mu^*,\kappa>0$, such that $u(t) \xrightarrow{\kappa \mu}[u_*^-,u_*^+]$ for all $\mu \in (0,\mu^*]$.
\hfill $\triangledown$
\end{lemma} 

\begin{proof}
The proof is similar to that of Theorem 1 in \cite{Angeli2014}.
Since the closed loop $u(t)$ is bounded in $[u^-(0),u^+(0)]:=[u^-_0,u^+_0]$, 
we have
\begin{align*}
    q(t) \xrightarrow{\mu} [\psi(u^-(0),u^+(0)),\psi(u^+(0),u^-(0))],
\end{align*}
By the cooperativity and Lipschitz property of $\Delta$, we have that $u(t) \to [u^-(1),u^+(1)]$, where
\begin{align*}
    u^-(1):&=\Delta \circ \psi(u^-(0),u^+(0))-L_\Delta \mu,\\
    u^+(1):&=\Delta \circ \psi(u^+(0),u^-(0))+L_\Delta \mu.
\end{align*}
After $(k+1)$-iterations, $u(t) \to [u^-(k+1),u^+(k+1)]$, where 
\begin{equation}
    \begin{aligned} \label{Eqn:DTIteration-pert}
    u^-(k+1) &=\Delta \circ \psi(u^-(k),u^+(k))-L_\Delta \mu,\\
    u^+(k+1) &= \Delta \circ \psi(u^+(k),u^-(k))+L_\Delta \mu.
\end{aligned}
\end{equation}
To study convergence of the this discrete time iteration, We treat it as a perturbation of the nominal system (\ref{Eqn:DTSys}). Since (\ref{Eqn:DTSys}) is exponentially ultimately bounded in $[u_*^-,u_*^+]$, we apply Lemma \ref{Lemma:RobustnessGUB} to prove ultimate boundedness of (\ref{Eqn:DTIteration-pert}). This provides a bound for the trajectory of the continuous time interconnected system because $u(t) \to [u^-(k),u^+(k)]$ for every integer $k \geq 0$.
\end{proof}

\noindent Since the singularly perturbed system (\ref{Eqn:RegulatedSubsystem}) has the approximate CICO property as shown in Lemma \ref{Lemma:SPCICS}, this small-gain theorem is directly applicable to study its feedback interconnection with a cooperative function $\Delta(\cdot)$. On the other hand, if the conditions for Lemma \ref{Lemma:SGT} are satisfied with $\mu=0$, then we have $u(t) \to [u_*^-,u_*^+]$. 

\subsection{Disturbance attenuation of feedback-regulated subsystems} \label{Sec:sRNA-DistAttn}
We show that $|h_i(r_i,0;\varepsilon_i)-r_i/\beta_i|$ and $|h_i(r_i,w_i;\varepsilon_i)-h_i(r_i,0;\varepsilon_i)|$ are both small in the following claims. Inequality (\ref{Eqn:sRNA.DistAttn}) can then be obtained via triangle inequality. 

\begin{claim} \label{Claim:hiSmooth}
\normalfont There exists $K^*_i>0$, independent of $r_i$, such that
\begin{align} \label{Eqn:sRNA-translationError}
    |h_i(r_i,0;\varepsilon_i)-r_i/\beta_i| \leq K^*_i \varepsilon_i
\end{align}
for all $r_i \in \bar{\mathcal{R}}_i$ and for $\varepsilon_i$ sufficiently small.
\hfill $\triangledown$
\end{claim}
\noindent The proof for a constant $r_i$ can be found in~\cite{Qian2018_JRSI}, and $K_i^*$ can be chosen independent of $r_i$ because $\bar{\mathcal{R}}_i$ is compact. 

\begin{claim} \label{Claim:Error2}
\normalfont Consider system (\ref{Eqn:sRNA-Reduced}), there exists a positive constant $k_i^*$, independent of $r_i$, such that for any fixed pair $(r_i,w_i) \in \bar{\mathcal{R}}_i \times \mathcal{W}_i$,
\begin{align} \label{Eqn:sRNA-translationError2}
    |h_i(r_i,w_i;\varepsilon_i)-h_i(r_i,0;\varepsilon_i)| \leq k^*_i \varepsilon_i |w_i| + K_i^* \varepsilon_i
\end{align}
for $\varepsilon_i$ sufficiently small, where $K_i^*$ is as defined in Claim \ref{Claim:hiSmooth}.
\end{claim}

\begin{proof}
To show Claim \ref{Claim:Error2}, we prove that $\limsup_{t \to \infty}|\bar{y}_i(t)-h_i(r_i,0;\varepsilon_i)| \leq \varepsilon_i k_i^*|w_i| + K_i^*\varepsilon_i$.
This is sufficient because we know $\bar{\Sigma}_i$ has a GAS equilibrium.
We first fix a $r_i \in \bar{\mathcal{R}}_i$, and let $y_i^*=h_i(r_i,0;\varepsilon_i)$ and $\tilde{y}_i:=\bar{y}_i-y_i^*$. 
The dynamics of $\tilde{y}_i$ follow:
\begin{align} \label{Eqn:sRNA-Reduced-2}
    \dot{\tilde{y}}_i= T_i(\tilde{y}_i,r_i,w_i) - \delta (y_i^* + \tilde{y}_i),
\end{align}
where
\begin{align*}
    T_i(\tilde{y}_i,r_i,w_i):=\alpha_i \frac{\bar{m}_i(\tilde{y}_i+y_i^*,r_i;\varepsilon_i)/\kappa_i}{1+\bar{m}_i(\tilde{y}_i+y_i^*,r_i;\varepsilon_i)/\kappa_i+w_i}.
\end{align*}
and because $T_i(0,r_i,0)- \delta y_i^* = 0$, we have $\bar{m}_i(y_i^*,r_i;\varepsilon_i) = \kappa_i \delta y_i^*/(\alpha_i-\delta y_i^*).$
Let $k_i(y_i^*):=\frac{\delta \kappa_i y_i^*}{\alpha_i-\delta y_i^*} \cdot \frac{2  \delta}{\beta_i}$,
we show that the trajectory of (\ref{Eqn:sRNA-Reduced-2}) is ultimately bounded in the set $\mathcal{P}_i(y_i^*):= \left\lbrace -k_i \varepsilon_i w_i-K_i^* \varepsilon_i \leq \tilde{y}_i \leq  0 \right\rbrace$ using the Lyapunov function $V_i(\tilde{y}_i) = \tilde{y}_i^2/2$. 
For $\tilde{y}_i \geq 0$, since $\partial T_i/\partial w_i, \partial T_i/\partial \tilde{y}_i<0$, we have $\dot{V}_i=\tilde{y}_i[T_i(\tilde{y}_i,w_i,r_i)-\delta x_i^*-\delta \tilde{y}_i] \leq  \tilde{y}_i[T_i(0,0,r_i)-\delta y_i^*-\delta \tilde{y}_i]=-2\delta V_i$. 
By Claim \ref{Claim:hiSmooth}, $y_i^* \leq r_i/\beta_i+K_i^*\varepsilon_i$, and therefore,
for $\tilde{y}_i \leq -k_i\varepsilon_iw_i-K_i^*\varepsilon_i <0$, we have
$\bar{y}_i=\tilde{y}_i+y_i^*\leq  r_i/\beta_i-k_i\varepsilon_iw_i$. 
We can use this to find that $\partial \bar{m}_i/\partial \tilde{y}_i \leq -\frac{\beta_i}{2\varepsilon_i\delta}$ for all $\tilde{y}_i \leq -k_i\varepsilon_iw_i-K_i^*\varepsilon_i$, and therefore $\bar{m}_i(y_i^*+\tilde{y}_i,r_i;\varepsilon_i) \geq \bar{m}_i(y_i^*,r_i;\varepsilon_i)(1+w_i)$ by mean value theorem.
Substituting into (\ref{Eqn:sRNA-Reduced-2}), we obtain
\begin{align*}
    T_i(\tilde{y}_i,r_i,w_i) \geq \alpha_i \frac{\gamma_i^1(y_i^*,r_i;\varepsilon_i)/\kappa_i}{1+\gamma_i^1(y_i^*,r_i;\varepsilon_i)/\kappa_i} = T_i(0,r_i,0) 
\end{align*}
if $\tilde{y}_i \leq -k_i\varepsilon_iw_i-K_i^*\varepsilon_i$. Thus, $\dot{V}_i=\tilde{y}_i[T_i(\tilde{y}_i,r_i,w_i)-\delta y_i^*-\delta \tilde{y}_i] \leq \tilde{y}_i[T_i(0,r_i,0)-\delta y_i^*-\delta \tilde{y}_i]=-2\delta V_i$.
Hence, we have shown that $\tilde{y}_i(t)$ eventually enters $\mathcal{P}_i$ for any fixed $(r_i,w_i) \in \bar{\mathcal{R}}_i \times \mathcal{W}_i$. Since $\bar{\mathcal{R}}_i$ is compact, due to Claim \ref{Claim:hiSmooth}, $y_i^*$ is also bounded in a compact set. Thus, there exists $k_i^* \geq k_i(y_i^*)$ for all $y_i^*$.
\end{proof}

\subsection{Lipschitz properties of subsystem characteristics} \label{Sec:GlobalLipschitz}

\noindent Since $\psi_i(r_i^+,w_i^+,r_i^-,w_i^-;\varepsilon_i)=\bar{\rho}_i(\bar{\varphi}_i(r_i^-,w_i^+;\varepsilon_i),r_i^+;\varepsilon_i)$, to show Assumption \ref{Ass:LipschitzConditions} is satisfied, we prove that $\bar{\rho}_i$ and $\bar{\varphi}_i=h_i$ each satisfies the Lipschitz conditions below.

\begin{claim} \label{Claim:sRNA-Lipschitz}
\normalfont There are positive functions $c_x(\cdot)$,  $c_r(\cdot)$ such that:
\begin{equation*} 
    \begin{aligned} 
    &|\bar{\rho}_i(p_i^+,r_i,w_i;\varepsilon_i)-\bar{\rho}_i(p_i^-,r_i,w_i;\varepsilon_i)| \leq c_p(\varepsilon_i)|p_i^+-p_i^-|,&\\
    &|\bar{\rho}_i(p_i,r_i^+,w_i;\varepsilon_i)-\bar{\rho}_i(p_i,r_i^-,w_i;\varepsilon_i)| \leq c_r(\varepsilon_i)|r_i^+-r_i^-|,&
\end{aligned}
\end{equation*}
$\forall (p_i,r_i,w_i;\varepsilon_i) \in \mathcal{X}_i \times \bar{\mathcal{R}}_i \times \mathcal{W}_i \times (0,\varepsilon_i^*]$.
In addition, $h_i(r_i,w_i;\varepsilon_i)$ is Lipschitz in $r_i \in \bar{\mathcal{R}}_i$ uniformly in $(w_i,\varepsilon_i) \in \mathcal{W}_i \times (0,\varepsilon_i^*]$.
\hfill $\triangledown$
\end{claim}

\begin{proof}
We first show the Lipschitz property of $h_i(r_i,w_i;\varepsilon_i)$. Since $\bar{\mathcal{R}}_i$ is an $\varepsilon_i$-independent compact subset of $(0,\alpha_i\beta_i/\delta)$, we let $\bar{\mathcal{R}}_i:=[\vartheta_i^1,\alpha_i\beta_i/\delta-\vartheta_2^i]$, where $0<\vartheta_1^i,\vartheta_2^i<\alpha_i \beta_i/\delta$ are $\varepsilon_i$-independent constants. 
Setting the dynamics of (\ref{Eqn:sRNA-Dyn-SP}) to steady state, the equilibrium $m_i$ is the solution to
\begin{align*}
\mathfrak{F}_i(m_i,r_i,w_i):=&\frac{\alpha_i \beta_i}{\delta} \frac{m_i/\kappa_i}{1+m_i/\kappa_i+w_i}-r_i\nonumber \\
+&\varepsilon_i \delta m_i- \varepsilon_i \frac{\delta r_i}{\lambda_i m_i} + \frac{\varepsilon_i^2 \delta}{\lambda_i} = 0,
\end{align*}
and the equilibrium output $y_i=p_i$ can be subsequently determined via 
\begin{align*}
    y_i=\mathfrak{G}_i(m_i,w_i)=\frac{\alpha_i}{\delta} \frac{m_i/\kappa_i}{1+m_i/\kappa_i+w_i}.
\end{align*}
Using chain rule and the implicit function theorem, we have $\frac{\partial h_i}{\partial r_i}=\frac{\partial \mathfrak{G}_i}{\partial m_i}\cdot \frac{\partial m_i}{\partial r_i}=-\frac{\partial \mathfrak{G}_i}{\partial m_i} \frac{\partial \mathfrak{F}_i}{\partial r_i} \left( \frac{\partial \mathfrak{F}_i}{\partial m_i}\right)^{-1},$
from which we find $0 < \frac{\partial h_i}{\partial r_i} \leq \frac{1}{\beta_i} + \frac{\alpha_i}{2 \delta \vartheta_i^1}$
for all $(r_i,w_i) \in \bar{\mathcal{R}}_i \times \mathcal{W}_i$. 
To show the Lipschitz properties of $\bar{\rho}_i$ are satisfied, we use (\ref{Eqn:sRNA.ControllerDoseResponse}) to find the following uniform bounds: $0<\frac{\partial \bar{m}_i}{\partial r_i} \leq \frac{1}{2\delta \varepsilon_i}$ and $-\frac{\beta_i}{2\delta \varepsilon_i} \leq \frac{\partial \bar{m}_i}{\partial p_i}<0.$
Since $\bar{\rho}_i=\bar{m}_i/\kappa_i$, we can take $c_r(\varepsilon_i)= \frac{1}{2\delta \min_i(\kappa_i) \varepsilon_i}$ and $c_p(\varepsilon_i)=\frac{\max_i(\beta_i)}{2\min_i(\kappa_i)\delta \varepsilon_i}$.
\end{proof}

\noindent Claim \ref{Claim:sRNA-Lipschitz} implies that
$\psi_i$ is Lipschitz in $(r_i^+,r_i^-) \in (\bar{\mathcal{R}}_i)^2$ uniformly in $w_i^-,w_i^+ \in \mathcal{W}_i$ with Lipschitz constant
$L_\psi(\varepsilon_i)=c_r(\varepsilon_i)+c_x(\varepsilon_i)L_h$.  The I/O gain function $\psi_i^*$ is sub-linear because, according to (\ref{Eqn:sRNA.IOG}), $\psi_i^*(w_i^+,w_i^-;r_i^*,\varepsilon_i)=\eta_i(r_i^*,w_i^+;\varepsilon_i)(1+w_i^+)$ and, as we have shown in the proof of Proposition \ref{Prop:sRNA}, $\eta_i$ is positive and bounded for $(r_i^*,w_i^+;\varepsilon_i) \in \bar{\mathcal{R}}_i \times \mathcal{W}_i \times (0,\varepsilon_i^*]$.

\bibliographystyle{unsrt}
{\footnotesize \bibliography{References_Yili}}


\end{document}